# Model of endogenous fluctuations of the norm of the wave function

**G. A. Ptitsyn**

*Semenov Institute of Chemical Physics*

*Russian Academy of Sciences*

*E-mail: gap19542017@gmail.com*

A theoretical model of endogenous fluctuations of the norm of the wave function, consistent with the standard quantum theory, is presented. These fluctuations are a subsystem of endogenous quantum fluctuations and describe one of the decoherence channels and the dynamics of the collapse of the quantum state.

## I. INTRODUCTION

Development of quantum technologies of computing, communication, teleportation, encryption, etc. bring to the fore the processes of loss of coherence and information in complex quantum systems, the end result of which is the collapse of the state. In this regard, the development of the theory of quantum decoherence and collapse is becoming increasingly relevant.

Standard quantum theory (SQT) does not describe the dynamics of the spontaneous catastrophic restructuring of the wave function (WF) that occurs during each measurement (and, therefore, when a quantum system interacts with any macroscopic system). This gap in the theory is partly due to the fact that the problem of "collapse" has been circumvented. Despite the fact that it is the device that destroys the wave function, it, paradoxically, does not affect the probability of obtaining a particular measurement result. These probabilities are completely determined by the view of the system WF before the measurement. Namely, the probability of obtaining the value of $a$ when measuring a physical quantity $A$ is determined by the expression:

$$w_A(a) = | < a|\psi > |^2, \tag{1}$$

where $|\psi >$ is the state vector of the system before measurement, and $< a|$ is the eigenvector of the operator of a physical quantity $A$, corresponding to the eigenvalue $a$. This universal law is called the "axiom of measurement".

We derive decoherence and collapse of the WF from the idea of universal and unremovable endogenous quantum fluctuations (EQF), to which, we believe, all particles of matter are susceptible [1–4]. In this case, the WF itself makes noise, and the total noise level is proportional to the number of particles in the physical system. In low-particle quantum systems, this noise is extremely small, and such systems evolve in accordance with the Schrödinger equation. However, when measured, such a system becomes entangled with a gigantic number of degrees of freedom of the macro-instrument, and the instrument noise leads to an almost instantaneous collapse of the combined WF of the system “particle + device”.

In the picture of the world that we offer, there is no fundamental difference between the “quantum particle” and the “classical device”. Everything is quantum here and everything is noisy. The difference is not qualitative, but quantitative - in terms of the level of self-noise. Few-particle systems behave as purely quantum ones. Macroscopic ones, on the contrary, are in a state of permanent collapse, which makes them macro-unambiguous, classical objects. Mesoscopic and few-particle systems at sufficiently large times demonstrate a gradual loss of coherence and information.

## II. MEASUREMENT AXIOM

From the above, an important conclusion can be made: to develop a theory of decoherence and collapse, there is no need to simulate a macro device and processes in it. It is enough to take one particle and choose for it such an algorithm of its own fluctuations that will satisfy the measurement axiom (1). Such fluctuations will be a constantly acting destructive factor in dynamics, which tends to destroy the quantum state and bring it to full collapse with the correct probabilities of final states. In the absence of an instrument, the particle seems to be trying to measure itself, i.e. localize itself in the phase space (more precisely, in the own representation of the full fluctuation operator [4]). But this is a long process, and it can be said that time plays the role of the instrument in this case.

If such a universal algorithm exists, then it will automatically ensure that the axiom of measurement is fulfilled even when the particle interacts with the macrodetector. The difference will be only in the speed of collapse. For this, the mechanism of fluctuations must ensure the cumulative effect of the noise of individual particles on the rate of destruction of the state. The postulated universality of the EQF algorithm immediately explains the above-mentioned independence of the collapse probabilities on the internal variables of the device.

In [4], the explicit form of the full EQF operator for a single particle and the algorithm of its action were already presented. However, there it was underdetermined in the region of small amplitudes of the WF. Our goal is to eliminate this flaw and find a general view of the algorithms that satisfy the measurement axiom.

To do this, we will simplify the problem in this study and build not a complete fluctuation operator, but only its part responsible for the destruction of wave packets and for the collapse of the state. To this end, we note that the complete fluctuation operator acts pointwise in phase space [4]. By its action, the operator slightly changes the norm of the wave packet in which the fluctuation occurred, as well as the shape of the WF in this packet. These two actions can be formally divided by presenting the fluctuation operator $\hat{F}$ as a product of two separate (already non-local) mutually commuting fluctuation operators of the norm $\hat{F}_n$ and the form $\hat{F}_f$:

$$\hat{F} = \hat{F}_n \hat{F}_f = \hat{F}_f \hat{F}_n.$$

The $\hat{F}_f$ operator adds impurities of new states to the WF, but does not change the WF norm, while $\hat{F}_n$, on the contrary, changes only the norm of one wave packet in the superposition of states, without changing its content at all. From this it follows that a set of arbitrary mutually orthogonal wave packets (in the representation of the complete fluctuation operator) (into which the WF of the particle can always be decomposed) can serve as the own representation of the norm fluctuation operator.

To simplify the development of the algorithm, we consider a particle not in an arbitrary, but in a specially prepared state consisting of $n$ spatially separated mutually orthogonal wave packets $|\varphi_i(t)>$, and put it in such conditions that these packets do not mix with each other. In this case, the packets will continue to be orthogonal to each other, although their shape will probably change. Since we are not currently interested in the form of the packages, we will temporarily replace the real evolving packages with abstract fixed orthonormal vectors $|i>$, which will play the role of a surrogate basis for $\hat{F}_n$. Surrogate WF has the form:

$$|\tilde{\psi}> = a_1|1> + a_2|2> + \cdots + a_n|n>, \quad (2)$$

$$\sum_{i=1}^{n}|a_i^2| = 1, \qquad <i|j> = \delta_{ij}, \qquad \delta_{ij} = \begin{cases} 1, & i = j \\ 0, & i \neq j \end{cases}.$$

If there were no proper fluctuations, then the norm of each packet $|a_i|$ persisted. In our model, we assume that the fluctuations of the wave function are repeated at small equal intervals of time τ, and the square of the norm of the wave packet during each fluctuation can change by a small but finite and fixed value ε≪1. Thus, the evolution of true WF is:

$$|\psi(t+\tau)> = \hat{F}\hat{U}(\tau)|\psi(t)> = \hat{F}_n\hat{F}_f\hat{U}(\tau)|\psi(t)>,$$

where $U(\tau)$ is the Schrödinger evolution operator. From the definition of $\hat{F}_n$ it follows that for WF, consisting of spatially separated wave packets, this operator will commute not only with $\hat{F}_f$, but also with $\hat{U}(\tau)$. And this means that fluctuations in the norm of packages can be considered completely separate from the evolution of the form of the packages themselves. Therefore, we will analyze the evolution of surrogate WF, which is much simpler:

$$|\tilde{\psi}(m\tau) >= \hat{F}_n{}^m|\tilde{\psi}(0) > .$$

The nonlinear random operator $\hat{F}_n$ does not change the basis vectors $|i >$, but acts, in fact, only on the coefficients of them. Further, we will work only with surrogate WF and therefore omit the "~" sign in its designation. The general view of true WF can be restored at any time by simply replacing the surrogate basis vectors $|i >$, with real wave packets changing under the influence of the evolution and form fluctuations operators:

$$|i > \xrightarrow{t=m\tau} |\varphi_i(m\tau) >= \left[\hat{F}_f\hat{U}(\tau)\right]^m|\varphi_i(0) > .$$

Consider one of the members of the superposition (2) and find out what conditions its fluctuation transitions must satisfy in order for its survival to satisfy the measurement axiom. For brevity, we will call the square of the norm of a package its "weight". If the packet weight is *x*:

$$|< i|\psi >|^2 = \left|a_i^2\right| = x, \qquad 0 \le x \le 1,$$

then, according to the measurement axiom, the probability $w_\infty(x)$ of the survival of this packet at $t \to \infty$ (in the absence of a macro device, time plays the role of a detector!) must be equal to *x*:

$$w_\infty(x) = x. \qquad (3)$$

The probability of a collapse of WF in this packet depends only on the current state of the quantum system and does not depend on its history. This means that the random process leading to collapse is Markovian.

If the total weight of all packets in the particle state is equal to one, then the probability of survival of the entire state, according to (3), is also equal to unity, i.e. one of the packages will surely survive:

$$w_\infty = w_\infty(x_1) + w_\infty(x_2) + \cdots w_\infty(x_n) = x_1 + x_2 + \cdots + x_n = 1. \qquad (4)$$

Thus, the axiom of measurement is self-consistent. But the condition of self-consistency (4) requires that the probability of a package's survival strictly corresponds to formula (3) for any *x*, even if it is arbitrarily small.

According to the model, with each fluctuation the weight of this package can either decrease ($x \to x - \varepsilon$), or remain unchanged ($x \to x$), or increase ($x \to x + \varepsilon$). Denote the probabilities of these transitions, respectively, $p(x),\ q(x),\ r(x)$. Thus, the fluctuations of the weight of each separate packet mathematically represent a wandering on the segment [0, 1] with absorbing ends, since the points $x = 0$ and $x = 1$ are stable for this random process (zero weight cannot increase, since this would mean vacuum instability, likewise, a unit weight cannot decrease, since it would mean the birth of a new package somewhere with a weight of up to unity, that is, vacuum instability again). For the ends of the segment we have:

$$w_\infty(0) = 0, \quad p(0) = r(0) = 0, \quad q(0) = 1;$$

$$w_\infty(1) = 1, \quad p(1) = r(1) = 0, \quad q(1) = 1. \tag{5}$$

The remaining points of the segment are unstable. The total probabilities of leaving an arbitrary point $x$ ($0 < x < 1$) down ($P(x)$), and up ($R(x)$) with regard to idle strokes are equal, obviously:

$$P = p + qp + q^2p + q^3p + \cdots = \frac{p}{1-q} = \frac{p}{p+r}, \tag{6}$$

$$R = \frac{r}{1-q} = \frac{r}{p+r}, \qquad P(x) + R(x) = 1.$$

The use of probabilities $P$ and $R$ allows one to get rid of taking into account the “misses” $q$, since the latter are effectively contained within them. According to the model, weights close to zero, $0 < x \leq \varepsilon$, may change by a fluctuation either to zero or to a point x + ε. As the fluctuations are a Markov process, so we have for them:

$$w_\infty(x) = P(x)w_\infty(0) + R(x)w_\infty(x + \varepsilon).$$

Substituting here (3), (6), we get:

$$x = R(x)(x + \varepsilon),$$

i.e.

$$R(x) = \frac{x}{x+\varepsilon}, \quad P(x) = \frac{\varepsilon}{x+\varepsilon}. \tag{7}$$

Similarly, for points close to unity, $1 - \varepsilon \leq x < 1$, two transitions are possible: to the point $x - \varepsilon$ and to the unity. Accordingly, we have:

$$w_\infty(x) = P(x)w_\infty(x - \varepsilon) + R(x)w_\infty(1).$$

Using (3) and (6), we get:

$$x = \big(1 - R(x)\big)(x - \varepsilon) + R(x) \cdot 1,$$

from where

$$R(x) = \frac{\varepsilon}{1-x+\varepsilon}, \;\; P(x) = \frac{1-x}{1-x+\varepsilon}. \qquad (8)$$

The probabilities (8) and (7) are not symmetric with respect to $P(x)$ and $R(x)$, but are symmetric with each other up to a simultaneous replacement of $x \leftrightarrow 1 - x$ and $P \leftrightarrow R$.

For all other points $\varepsilon \leq x \leq 1 - \varepsilon$, symmetric transitions to the step ε up and down are allowed. For them we have:

$$w_\infty(x) = P(x) w_\infty(x - \varepsilon) + R(x) w_\infty(x + \varepsilon).$$

Substituting here (3) and (6), we get:

$$x = (1 - R(x))(x - \varepsilon) + R(x)(x + \varepsilon),$$

from where

$$P(x) = R(x) = 1/2. \qquad (9)$$

Combining (5) - (9), we finally get for all $x$:

$$\begin{cases} p(0) = r(0) = 0, \quad q\,(0) = 1; & x = 0 \\ p(x) = \dfrac{\varepsilon[1 - q(x)]}{x + \varepsilon}, \; r(x) = \dfrac{x[1 - q(x)]}{x + \varepsilon}; & 0 \leq x \leq \varepsilon \\ p(x) = r(x) = \dfrac{1 - q(x)}{2}; & \varepsilon \leq x \leq 1 - \varepsilon \\ p(x) = \dfrac{(1 - x)[1 - q(x)]}{1 - x + \varepsilon}, \;\; r(x) = \dfrac{\varepsilon[1 - q(x)]}{1 - x + \varepsilon}; & 1 - \varepsilon \leq x \leq 1 \\ p(1) = r(1) = 0, \quad q(1) = 1; & x = 1 \end{cases}$$

$$w_\infty(x) = x; \quad 0 \leq x \leq 1. \qquad (10)$$

The constructed model of fluctuation transitions (10) strictly satisfies the measurement axiom for any packet weight. Moreover, it contains an arbitrary for the time being (except for the end points) function $q(x)$ - probabilities of “misses”. “Misses” do not affect the probability of survival and destruction of packages. But their share, of course, will affect the kinetics of decoherence of quantum ensembles and the rate of quantum selection.

In fact, the function $q(x)$ is not arbitrary, since in addition to the measurement axiom, there are several other criteria for compliance with the standard quantum theory, which we will discuss below.

## III. COLLAPSE ADDITIVITY

Scheme (10) is not a ready-made fluctuation algorithm, since it sets the transition probabilities for only one wave packet from the full superposition (2). To obtain an algorithm for the entire WF, one must take into account the connections that exist between the individual state packets. Namely, at each tick of the dynamics with duration τ:

- the full norm of WF is conserved;
- exactly one fluctuation occurs in the whole WF.

To account for connections, you must include the entire WF in the scheme. The conservation of its norm means that in each fluctuation the weight loss in one package is compensated by the profit in some other. That is, each fluctuation contains two independent actions — negative and positive semi-fluctuations (NSF and PSF), and the fluctuation operator of the norm $F_n$ splits into the product of two corresponding operators $\hat{F}_{n-}$, $\hat{F}_{n+}$.

$$\hat{F}|\psi> = \hat{F}_{+}\,\hat{F}_{-}|\psi>. \tag{11}$$

(we omit the index n of these operators, since we will deal only with the fluctuations of the norm).

Random operators $\hat{F}_{+},\ \hat{F}_{-}$ - are not commutative. In particular, the NSF operator destroys packets of light weight, after which the PSF in the destroyed packet is no longer possible, in contrast to the reverse procedure. We accept the order in which the NSF occurs first, and then the PSF, as written in (11). It is important to note that the operators $\hat{F}_{+},\ \hat{F}_{-}$ denote the internal non-linear rearrangement of WF and therefore are unidirectional. In expressions like $< a\left|\hat{F}_{\pm}\right|\psi >$, they act only to the right and only on the ket-vector of the state. Unlike linear operators of physical quantities, their action cannot be redirected to the left — to the representation vector, whatever that representation may be. A better record would be: $< a|\hat{F}_{\pm}\psi >$, but it is not always convenient. Hermitian conjugate to $\hat{F}_{+}, \hat{F}_{-}$ operators act, on the contrary, only to the left and only on the bra-state vector.

In the experiment, the probability of registration of a particle is proportional to the sum of the weights of the wave packets trapped in the detector. It is easy to prove that to fulfill this property of "additivity of a collapse" it is enough that the algorithms of each of the two semi-fluctuations are additive (in the weight of the packets). And for this, the probability of semi-fluctuations for two arbitrary packages, formally considered as one combined package, should be equal to the sum of probabilities of semi-fluctuations of each of the packages separately, i.e.:

$$p_{\mp}(x_1 + x_2) = p_{\mp}(x_1) + p_{\mp}(x_2).$$

This condition is satisfied if the probabilities are proportional to the weights of the packets. We should normalize them at the (current) total weight of WF:

$$p_{i\,\mp} = \frac{x_i}{<\psi|\psi>} = \frac{|<i|\psi>|^2}{<\psi|\psi>}. \quad (12)$$

In this case, the probability of each semi-fluctuation over the entire WF is equal to one, i.e. it will definitely happen somewhere, and the scheme is closed.

Now we will build anew the transition scheme for a single wave packet, taking into account the additivity of the semi-fluctuations.

The weight $x$ of the wave packet will decrease if the NSF occurs in this packet, and the PSF in some another one. The weight of the package will increase with the opposite localization of events. The probabilities $p(x),\ r(x)$ of these pairs of independent events are equal to the product of their probabilities. According to (12) it gives:

$$\begin{cases} p(x) = r(x) = \frac{x(1-x)}{(1-\varepsilon)} \\ q(x) = 1 - p(x) - r(x) = 1 - \frac{2x(1-x)}{(1-\varepsilon)} \end{cases}, \qquad \varepsilon \le x \le 1 - \varepsilon. \quad (13)$$

Here $(1-\varepsilon)$ is the weight of the entire state after the NSF.

The comparison shows that scheme (13) is a representative of the class of schemes (10) in the range of package weights ($\varepsilon \leq x \leq 1-\varepsilon$). This means that in this range it strictly satisfies the measurement axiom. The sides of the weight range ($0 < x < \varepsilon,\ 1-\varepsilon < x < 1$) require separate consideration. The problem with small weights ($x < \varepsilon$) is that when destroying such a package as a result of NSF, the weight loss of the state does not reach the planned value of ε, and this does not allow to satisfy all the criteria of compliance with SQT and, in particular, the measurement axiom. This means that the fluctuation algorithm for small packets needs to be determined. One possibility is the cascade NSF. In this case, NSF is redefined not as a single event, but as a series of consecutive draws according to the standard NSF rules (12), which are repeated until the total weight loss reaches exactly ε. After the end of the cascade, exactly one PSF is produced according to the standard scheme with the restoration of the WF weight to one. The rationale for this recipe will be given below.

For now, let us consider the relaxation kinetics of a statistical ensemble of wave packets, whose weight is multiple to the fluctuation step ε: $x = m\varepsilon$ (each package also has some other packages that complement the total weight of each state to one, but now we will select only one representative from each state).

According to the scheme, fluctuation transitions here go only at points of the same set $x = m\varepsilon$. We write the current state of this statistical ensemble in one column vector:

$$|\Phi> = (n_0, n_1, n_2, \dots, n_M)^T .$$

Here $n_m$ is the share of packets in the ensemble with weight $m\varepsilon$; $M = 1/\varepsilon$ is a large integer; T - transpose symbol. With such a record, the dynamics of fluctuation transitions in an ensemble can be described by a three-diagonal statistical matrix $S$:

$$|\Phi(t+\tau)> = S|\Phi(t)> ,$$

$$S = \begin{pmatrix} 1 & p_1 & & & \cdots & & & \\ & q_1 & p_2 & & \cdots & & & \\ & r_1 & q_2 & p_3 & \cdots & & & \\ & & r_2 & q_3 & \cdots & & & \\ \vdots & \vdots & \vdots & \vdots & \ddots & \vdots & \vdots & \vdots \\ & & & & \cdots & q_{M-2} & p_{M-1} & \\ & & & & \cdots & r_{M-2} & q_{M-1} & \\ & & & & \cdots & & r_{M-1} & 1 \end{pmatrix}.$$

Here $p_m, q_m, r_m \equiv p(m\varepsilon), q(m\varepsilon), r(m\varepsilon)$ from (13). Since $S$ is a constant matrix, the evolution of the ensemble of the packets is described by a simple equation:

$$|\Phi(n\tau) >= S^n |\Phi(0) >.$$

The solution to this equation is:

$$|\Phi(n\tau) >= \sum_{k=0}^{M} |R_k > C_k {\lambda_k}^n, \qquad C_k = < L_k|\Phi(0) >,$$

$$\lambda_k = 1 - \frac{\varepsilon^2 k(k-1)}{1-\varepsilon}, \qquad k = 0, 1, 2, \dots, M.$$

Here, $\lambda_k$ is the eigenvalues of the matrix $S$; $< L_k|$, $|R_k >$ are respectively its left and right eigenvectors with the properties:

$$< L_k|S = \lambda_k < L_k|, \qquad S|R_k >= \lambda_k|R_k >,$$

$$< R_k|R_k >= 1, \qquad < L_k|R_{k'} >= \delta_{kk'}.$$

The maximum eigenvalues $\lambda_0 = \lambda_1 = 1$ correspond to two stable states — the zero and unit weights of all packets of the ensemble. The remaining ($M$ - 1) eigenvalues are distributed in the range [0, 1), ensuring the relaxation of the ensemble with a large (for $M = \frac{1}{\varepsilon} \gg 1$) set of relaxation times $T_k = -\tau/ln\lambda_k$. The greatest of these times:

$$T_2 = -\frac{\tau}{ln\lambda_2} \cong \frac{\tau}{2\varepsilon^2},$$

can be called "the time of quantum selection." During this characteristic time, endogenous fluctuations completely destroy all wave packets of the initial state of an isolated particle except only random one, which as a result will survive and gain maximum weight due to the lost brethren.

Estimates of the possible values of the parameters of the model ε, τ can be given only after the generalization of the algorithm of fluctuations to many-particle systems and the consideration of model decoherence and measurement processes.

The constructed model of fluctuation transitions is not yet a ready-made algorithm. There are some more criteria for compliance with the standard quantum theory, which need to be satisfied before starting computer simulations of fluctuation quantum dynamics.

## IV. AVERAGE VALUES OF OBSERVABLES

To formulate the following criteria for SQT compliance, we need an explicit representation of the semi-fluctuation operators. According to the scheme, the NSF operator acts on a random wave packet $|k>$ in superposition (2) and, if possible, reduces its weight $x_k$ by a fixed value $\varepsilon \ll 1$, or completely destroys the packet if its weight does not reach ε. Denote the WF weight loss with a single NSF draw through $\epsilon$:

$$\epsilon = \begin{cases} \varepsilon, & x_k \geq \varepsilon \\ x_k, & x_k < \varepsilon \end{cases}, \qquad x_k = |< k|\psi > |^2 = |a_k|^2. \tag{14}$$

Then we get the following expression for the NSF operator:

$$\hat{F}_-|\psi> = (1 + |k > n_k < k|)|\psi>, \tag{15}$$

where $n_k$ is a numerical factor whose value is determined by the required reduction in the weight of WF as a result of NSF:

$$< \psi|\hat{F}_-^\dagger \hat{F}_-|\psi > = 1 - \epsilon, \qquad < \psi|\psi > = 1.$$

Substituting here (15), we get:

$$1 - \epsilon = < \psi|F_-^\dagger F_-|\psi > = < \psi|[1 + |k > (n_k + n_k^* + |n_k^2|) < k|]|\psi > =$$

$$= < \psi|\,\psi > + (-1 + |1 + n_k|^2)\;| < k|\psi > |^2,$$

i.e.

$$1 - \epsilon = 1 + (-1 + |1 + n_k|^2)x_k. \tag{16}$$

Solving this algebraic equation taking into account (14), we get:

$$n_k = -1 + e^{i\xi}\sqrt{1 - \frac{\epsilon}{x_k}} = \begin{cases} -1 + e^{i\xi}\sqrt{1 - \varepsilon/x_k}, & x_k \geq \varepsilon \\ -1, & x_k \leq \varepsilon \end{cases}, \qquad (17)$$

where $e^{i\xi}$ is an arbitrary (so far) phase factor.

Positive semi-fluctuation is easier than negative one. There are no cascades here and there is always an increase in the weight of a random packet exactly on ε. The PSF operator is:

$$\hat{F}_+|\psi' > = (1 + |k > p_k < k|)|\psi' >, \qquad (18)$$

where $|\psi' > = \hat{F}_-|\psi >$ - WF with reduced weight after NSF, and $p_k$ is a numerical factor whose value is determined by the required increase in weight of WF as a result of PSF:

$$< \psi'|\hat{F}_+^\dagger \hat{F}_+|\psi' > = 1, \qquad < \psi'|\psi' > = 1 - \varepsilon. \qquad (19)$$

Substituting here (18) and solving a similar to (16) algebraic equation, we get:

$$p_k = -1 + e^{i\eta}\sqrt{1 + \frac{\varepsilon}{x'_k}}, \qquad x'_k = |< k|\psi' >|^2, \qquad (20)$$

where $e^{i\eta}$ is an arbitrary phase factor.

To find the phase factors in (17), (20), it is necessary to use another criterion of the model conformity to the standard quantum theory. This is the requirement of equality of the average values of physical quantities obtained in the fluctuation model and in SQT, in accordance with the standard statistical interpretation of the wave function:

$$\bar{A}(t) = Sp[\bar{\varrho}(t)\hat{A}] = Sp[\varrho_S(t)\hat{A}].$$

Here $\bar{\varrho}(t)$ is the density matrix of the fluctuating system, averaged over the statistical ensemble; $\varrho_S(t)$ is the SQT density matrix with the same initial conditions; $\hat{A}, \bar{A}(t)$ are the operator of an arbitrary observable and its mean over the quantum ensemble. The solution of this equation with an arbitrary Hermitian operator $\hat{A}$ for WF (2) leads to a system of equalities for the fluctuating and standard WFs:

$$\begin{cases} \overline{|< i|\psi(t) > |^2} = |< i|\psi_S(t) >|^2 \\ \overline{< i|\psi(t) >} = < i|\psi_S(t) > \end{cases}, \qquad i = 1, 2, \ldots, n. \qquad (21)$$

In order for these equalities to be fulfilled at an arbitrary point in time, they must be satisfied at each step of the dynamics, namely, before and after the regular fluctuation. A sufficient condition for this is the implementation of similar proportionality for both operators of semi-fluctuations separately:

$$\begin{cases} \overline{\left|< \imath|\hat{F}_{\mp}|\psi(t) > \right|^2} = k_{\mp}|< i|\psi(t) >|^2 \\ \overline{< \imath|\hat{F}_{\mp}|\psi(t) >} = c_{\mp} < i|\psi(t) > \end{cases}, \qquad \begin{matrix} i = 1, 2, \dots, n; \\ k_{\mp},\ c_{\mp} = const. \end{matrix} \qquad \begin{matrix} (22) \\ (23) \end{matrix}$$

### A. Negative semi-fluctuation

We start with equation (22) for the NSF. Before averaging the left side, we have:

$$\left|< i|\hat{F}_{-}|\psi(t) > \right|^2 = |< i|\ (1 + |k > n_k < k|\ )|\psi >|^2 =$$
$$= |(1 + \ \delta_{ik} n_i) < i|\psi > |^2.$$

The probability of the NSF event dropping to the $k$-th packet is equal to:

$$w_{k-} = \frac{x_k}{<\psi|\psi>} = x_k = |< i|\psi(t) >|^2. \qquad (24)$$

Averaging over the ensemble with regard to (17), (24) gives:

$$\overline{\left|< \imath|\hat{F}_{-}|\psi(t) > \right|^2} = (1 - \varepsilon)|< i|\psi(t) >|^2, \qquad x_k \geq \varepsilon. \qquad (25)$$

Thus, equality (22) for NSF is exactly fulfilled for all packages with weights exceeding ε. For the constant, we get the value

$$k_{-} = (1 - \ \varepsilon).$$

For packages of lower weight, the cascade NSF procedure is applied. Namely, if the first NSF falls on a packet of small weight $x_1 < \varepsilon$, then this packet is destroyed, the NSF procedure repeats according to the same rules, but with probabilities (24), corrected for the decreased weight of the state:

$$w_{k-} = x_k/(1 - x_1).$$

These procedures with the destruction of packages are repeated until the complete loss of weight reaches exactly ε (that is, the last packet in the cascade is not destroyed, but only loses part of its weight). In the addendum to the main text, we show that after such a cascade, equality (25) is already satisfied for small packets (with deviations that have the next order of smallness in the fluctuation step ε):

$$\overline{\left|< \imath|\hat{F}_{-}|\psi(t) >\right|^2} = (1-\varepsilon)|< i|\psi(t) >|^2 + \Delta_k,$$

$$\Delta_k \le \varepsilon^2 \frac{x_k}{\varepsilon}\left(1-\frac{x_k}{\varepsilon}\right) \le \frac{\varepsilon^2}{4}, \qquad x_k < \varepsilon.$$

In addition, in the theory of the full operator of fluctuations (and not the operator of fluctuations of the norm discussed here) there arise other possibilities for combining small amplitudes that do not require repeated NSF drawings. And for them, equality (25) holds exactly in the whole space.

We now turn to equality (23) for the NSF. Opening the left side, we get:

$$\overline{< \imath|\hat{F}_{-}|\psi >} = < i|\left(1 + \overline{|k > n_k < k|}\right)|\psi > = (1+\overline{n}_\imath x_i) < i|\psi >.$$

Combining this with the right-hand side of (23), we get:

$$\overline{n_\imath} = \frac{c_- - 1}{x_i} = -\frac{c\prime}{x_i}, \qquad c_- = 1 - c'. \tag{26}$$

For large packages ($x_k \ge \varepsilon$), substituting here the value of $n_i$ from (17), we have:

$$\overline{e^{\imath\xi}} \equiv \theta_- = \frac{1-c\prime/x_i}{\sqrt{1-\varepsilon/x_i}}, \qquad \varepsilon \le x_i < 1\,.$$

Since $|\theta_-| \le 1$, the only possible value of the constant $c'$ in the whole range $\varepsilon \le x_i < 1$ is $c' = \varepsilon$. Finally, we get:

$$\theta_- = \overline{e^{\imath\xi}} = \sqrt{1-\varepsilon/x_i}\,, \qquad c_- = 1-\varepsilon\,, \qquad \varepsilon \le x_i < 1\,. \tag{27}$$

Rewriting (23), we get:

$$\overline{< \imath|\hat{F}_{-}|\psi >} = (1-\varepsilon) < i|\psi >\,, \qquad \varepsilon \le x_i < 1\,.$$

Condition (27) can be satisfied in many ways. For example:

$$e^{i\xi} = \begin{cases} 1, & p = \sqrt{1-\varepsilon/x_i} \\ \pm i, & q = (1-p)/2 \end{cases}, \qquad \varepsilon \le x_i < 1\,. \tag{28}$$

($p$, $q$ - probabilities of events). Thus, for large packets, equality (27) is strictly fulfilled. The distribution function of the phase factor (28) is chosen arbitrarily. There are no criteria except (27) for its selection yet.

For small packets ($x_k \leq \varepsilon$), substituting the value $n_i = -1$ from (17) into (26), we have:

$$c_{i-} = 1 - x_i, \qquad x_k < \varepsilon. \qquad (29)$$

Thus, equality (23) is slightly violated for small packets. The relative magnitude of the violation is of the order of $\varepsilon \ll 1$.

## B. Positive semi-fluctuation

PSF is simpler than NSF. The total weight of the state after NSF is $(1-\varepsilon)$, so nothing prevents us to increase the weight of any packet by exactly ε. Find the constant $k_+$ from (22). Opening the left part with the help of (18) and averaging it taking into account the normalization (19), we get:

$$\overline{\left|< \imath\left|\hat{F}_+\right|\psi' >\right|^2} = |< i|\psi' > |^2 \sum_{k=1}^{n} w_{k+}|1 + \ \delta_{ik}p_k|^2 =$$

$$= |< i|\psi' > |^2 \left[1 + (-1 + |1 + p_i|^2)\frac{x_i'}{1-\varepsilon}\right] = \frac{|<i|\psi\prime>|^2}{1- \varepsilon}. \qquad (30)$$

In this way,

$$k_+ = 1/(1- \ \varepsilon)$$

for all $i$. Recalling now (19) that $|\psi' > = \ \hat{F}_-|\psi >$, and using (25), (30), we obtain the averaging over both semi-fluctuations:

$$\overline{\left|< \imath\left|\hat{F}_+\hat{F}_-\right|\psi(t) >\right|^2} \ = \frac{\overline{|<\imath|\hat{F}_-|\psi(t)>|^2}}{1- \varepsilon} = |< i|\psi(t) >|^2 + O(\varepsilon^2) \,.$$

This equality is valid (with accuracy $\varepsilon^2/4$) in the whole space.

Finally, we consider equality (23) for PSF. Opening the left side, we get:

$$\overline{< \imath\left|\hat{F}_+\right|\psi' >} \ = < i\left|\left(1 + \left|\overline{k > p_k < k}\right|\right)\right|\psi' > = \ (1 + \overline{p_\imath}x'_i) < i|\psi' >.$$

Combining this with the right-hand side of (23), we get:

$$\overline{p_\imath} = \frac{c_+ - 1}{x'_i} = \ -\frac{c\prime}{x'_i} \,, \qquad c_+ = 1 - c'.$$

Substituting the value of $p_i$ from (20), we obtain:

$$\overline{e^{\imath\eta}} = \theta_+ = \frac{1 + c'/x'_i}{\sqrt{1+\varepsilon/x'_i}}, \qquad 0 < x'_i < 1 \,. \qquad (31)$$

Since $|\theta_+| \le 1$, then in this case the only possible value of the constant for all $x'_i$ is $c' = 0$. Finally, we get:

$$\theta_+ = \left(1 + \frac{\varepsilon}{x'_i}\right)^{-1/2}, \qquad c_+ = 1, \qquad 0 < x'_i < 1\,.$$

Condition (31) can be met in many ways. For example:

$$e^{i\eta} = \begin{cases} 1, & p = 1/\sqrt{1 + \varepsilon/x'_i} \\ \pm i, & q = (1-p)/2 \end{cases}, \qquad 0 < x'_i < 1\,. \tag{32}$$

There are no theoretical criteria for choosing the distribution of this phase factor. It is possible that some information about the distributions (28) and (32) can be gleaned by studying decoherence in interference experiments.

Rewriting (23), we get:

$$\overline{< \imath|\hat{F}_+|\psi' >} = < i|\psi' >, \qquad 0 < x'_i < 1\,.$$

Averaging over both semi-fluctuations, we get:

$$\overline{< \imath|\hat{F}_+\hat{F}_-|\psi >} = (1-\epsilon) < i|\psi >, \quad \epsilon = \min(\varepsilon, x_i),$$

which is the best possible approximation of the desired equality (21), given that $\epsilon \ll 1$.

Let us collect, at the end, the complete EQF algorithm of our simplified fluctuation model, returning to the original notation. So, we are dealing with a single-particle WF that consists of mutually orthogonal spatially separated wave packets.

$$|\psi > = a_1|\varphi_1 > + a_2|\varphi_2 > + \cdots + a_n|\varphi_n >,$$

$$< \varphi_i|\varphi_j > = \delta_{ij}, \qquad \sum_{i=1}^{n}|a_i^2| = 1.$$

At equal microscopic time intervals τ, the endogenous fluctuation described by the random nonlinear operator $\hat{F}$ occurs in the wave function of the particle. In the intervals between fluctuations WF evolves under the control of the standard Schrödinger evolution operator $\hat{U}(t)$. The fluctuation operator can be artificially divided into the product of the fluctuation operators of the norm $F_n$ and the form $F_f$:

$$\hat{F} = \hat{F}_n\hat{F}_f = \hat{F}_f\hat{F}_n.$$

Evolution in one dynamic cycle τ is described by the equation:

$$\left|\psi(t+\tau)>=\hat{F}\hat{U}(\tau)\right|\psi(t)>=\hat{F}_n\hat{F}_f\hat{U}(\tau)|\psi(t)>,$$

and, for the type of WF chosen by us, the combination of the operators $\hat{F}_f\hat{U}(t)$ acts only on the wave packets $|\varphi_i>$, changing their form, and the operator $\hat{F}_n$ affects, in fact, only the coefficients $a_i$ for them. Thus, the dynamics are split into two parallel and independent processes: changes in the shape of wave packets and fluctuations of their norms.

Leaving aside the change in the shape of the packets, we constructed the explicit form of only the fluctuation operator of the norm $\hat{F}_n$. This operator splits into the product of non-commuting operators of negative $\hat{F}_{n-}$ and positive $\hat{F}_{n+}$ semifluctuations.

$$\hat{F}_n=\hat{F}_{n+}\hat{F}_{n-}$$

The first is a negative semi-fluctuation in a random packet $|\varphi_k>$ among those available with probability

$$w_{k-}=\frac{x_k}{<\psi(t)|\psi(t)>},\qquad x_k=|<i|\psi(t)>|^2\,, \tag{33}$$

where $<\psi(t)|\psi(t)>$ is the total weight of WF before this NSF. Weight of this packet decreases by ε as a result of the NSF, and if the weight of the packet does not reach ε, then it is completely destroyed, and the NSF procedure repeats according to the same scenario, but with probabilities (33) adjusted due to the slightly changed WF weight. The NSF repetition cascade continues until the total weight loss of WF reaches exactly ε. If the package is not completely destroyed as a result of NSF, then not only its weight, but also the complex phase changes. The explicit form of the NSF operator is:

$$\hat{F}_-|\psi>=(1+|k>n_k<k|)|\psi>,$$

$$n_k=\begin{cases}-1+e^{i\xi}\sqrt{1-\varepsilon'/x_k}, & x_k\geq\varepsilon'\\ -1, & x_k\leq\varepsilon'\end{cases},$$

$$\overline{e^{i\xi}}=\sqrt{1-\varepsilon'/x_k},\qquad\qquad \varepsilon\leq x_k<1,$$

$$e^{i\xi}=\begin{cases}1, & p=\sqrt{1-\varepsilon'/x_k}\\ \pm i, & q=(1-p)/2\end{cases}.$$

Here $p$, $q$ are the probabilities of events, and $\varepsilon' = \varepsilon$ in the case of a single NSF, and $\varepsilon' < \varepsilon$ in the case of a cascade NSF.

A positive semi-fluctuation follows - also in a random packet with the same functional probability.

$$w_{k+} = \frac{x'_k}{< \psi'|\psi' >}. \qquad x'_k = |< i|\psi' >|^2,$$

where $|\psi' >$, $x'_k$ are WF and the packet weight modified by the preceding NSF. As a result of the PSF, weight of one packet increases exactly by ε, thereby restoring the total weight of WF to one. In addition to weight, its phase also changes. The clear view of the PSF operator is:

$$\hat{F}_+|\psi' > = (1 + |k > p_k < k|)|\psi' >,$$

$$p_k = -1 + e^{i\eta}\sqrt{1 + \frac{\varepsilon}{x'_k}}, \qquad\qquad x'_k = |< k|\psi' >|^2,$$

$$\overline{e^{i\eta}} = 1/\sqrt{1 + \varepsilon/x'_k}, \qquad e^{i\eta} = \begin{cases} 1, & p = 1/\sqrt{1 + \varepsilon/x'_k} \\ \pm i, & q = (1 - p)/2 \end{cases}.$$

## V. CONCLUSION

The developed algorithm for the fluctuations of the norm will be a component part of the algorithm of the full operator of fluctuations. However, it is interesting in itself. Due to its simplicity, it can be used to accelerate computing instead of the full algorithm of fluctuations in cases where the wave function of a particle consists of a large number of wave packets and when the exact shape of these packets is not important for some reason. In this case surrogate packets that obey the standard Schrödinger equation can be used in simulations instead of real wave packets, i.e. one may neglect the fluctuations of their form and leave in the dynamics only the fluctuations of their norm.

The fluctuation approach to quantum theory that we are developing provides a justification for the statistical interpretation of the wave function in standard quantum theory, explicitly introducing into the dynamics a random process of endogenous fluctuations responsible for the indeterministic result of quantum measurements and for the loss of coherence and information in quantum systems. However, unlike the standard wave function, which has the mathematical status of a "probability wave", the fluctuating wave function in our picture of the world has the physical status of a real noisy self-action field, but it is completely unobservable. An observable is a function averaged over a quantum ensemble, which coincides with a standard one, which can be measured by the method of quantum tomography, but not a single realization of a random wave function.

## VI. ADDENDUM. CASCADE SEMI-FLUCTUATION

Here we will analyze the cascade negative semi-fluctuation algorithm. We decided to move it to the end of the article in order to clarify the logic of the main text. However, this part of the algorithm is one of the main goals of the article, since it closes the fluctuation algorithm when handling wave packets of small weight, and it was this part that was not ready in the previously published version of the algorithm [4].

According to the scheme, each endogenous fluctuation begins with a negative semi-fluctuation. In the superposition of states (2)

$$|\tilde{\psi}> = a_1|1> + a_2|2> + \cdots + a_n|n>, \tag{34}$$

$$<i|j> = \delta_{ij}, \qquad \sum_{i=1}^{n}\left|a_i^2\right| = 1,$$

the weight of one of the wave packets decreases by ε as a result of NSF. If the weight of the package is less than ε, then the package is completely destroyed. But here a problem arises, because the symmetric probabilities of positive and negative semi-fluctuations (12):

$$p_{i\mp} = \frac{x_i}{<\psi|\psi>} = \frac{|<i|\psi>|^2}{<\psi|\psi>}, \tag{35}$$

dictated by the principle of "additivity of the collapse", it is impossible to reconcile with the asymmetric transition probabilities dictated for small packets by the measurement axiom (10):

$$p(x) = \frac{\varepsilon[1-q(x)]}{x+\varepsilon}, \qquad r(x) = \frac{x[1-q(x)]}{x+\varepsilon}, \qquad 0 \le x \le \varepsilon.$$

Reconciliation is possible (and was carried out above) for packages "heavier" than ε only. Thus, the algorithm needs to be modified for lightweight packages.

The strict requirements of the model are as follows:

- the WF norm should be conserved after each fluctuation (NSF + PSF);
- small packages must die (otherwise there will be no collapse);
- the collapse should be additive for any packages;
- the survival of any package must comply with the axiom of measurement;

A less stringent requirement: the average values of the physical quantities obtained in the fluctuation model and in the standard quantum theory must coincide.

To meet strict requirements, the weight loss in the NSF must be brought to the standard value of ε. One way to achieve this is to cascade repetitive NSFs. After the destruction of a package with weight $x_1 < \varepsilon$, a repeated drawing is carried out according to the same NSF rules, but with probabilities (35), corrected for the decreased weight of the state $< \psi|\psi > = 1 - x_1$. If the second NSF again falls into a small packet and $x_2 < \varepsilon - x_1$, then this packet is also destroyed, and the third rally is held, etc. The last package in the series is not destroyed, but only decreases in weight so that the total weight loss is ε. After that, the only PSF is carried out according to standard rules.

This algorithm saves the WF norm and destroys small packets. It is additive in PSF. It is additive in NSF as well, as each draw in the cascade is additive. It satisfies the measurement axiom, since the weight gain rules for PSF did not change, and the NSF, as required, reduces the weight of WF by exactly ε, simply using several random packets if one is not enough. As a result of the cascade addition of the weights of small packets, the algorithm never falls into the small weights area, where the requirements of the two strict principles do not agree with each other. Thus, all the stringent requirements of the cascade NSF model are met. It remains for us to check the behaviour of the average values.

It was shown above that the requirement of equality of average values in the model and SQT lead to a system of equalities that must be satisfied separately for the NSF and PSF. The PSF algorithm has not been changed, so the equalities obtained there will be fulfilled here. And in relation to the NSF, they are formulated as follows:

$$\begin{cases} \overline{\left|< \imath|\hat{F}_{-}|\psi(t) >\right|^2} = k_{-}|< i|\psi(t) >|^2 \\ \overline{< \imath|\hat{F}_{-}|\psi(t) >} \ = c_{-} < i|\psi(t) > \end{cases}, \quad \begin{matrix} i = 1, 2, \ldots, n; & (36) \\ k_{-} = c_{-} = 1 - \varepsilon. & (37) \end{matrix}$$

Here $\hat{F}_{-}$ is the cascade NSF operator, averaging goes over the quantum ensemble, and equalities must be fulfilled for each cascade.

If there are small packets in WF, then there will be cascades when implementing NSF. However, the length of the cascade itself is a random value. Because of this, it is not possible to calculate the average values in general. Consider special cases. We start with equalities (36).

Let WF consist of one small wave packet and a number of large ones, which we formally combine into one combined packet, which will facilitate the analysis. (This is permissible due to the additivity of fluctuation schemes in weights. And we are not interested in how NSFs are distributed over large packets, since the NSF algorithm is debugged for them, and there can be no surprises.) We write the weights of the two formed packages in the form of a single column vector::

$$\begin{pmatrix} |< 1|\psi >|^2 \\ |< 2|\psi >|^2 \end{pmatrix} = \begin{pmatrix} x \\ y \end{pmatrix}, \qquad x + y = 1, \qquad x < \varepsilon.$$

After the first NSF, we obtain, in accordance with the algorithm and the expression for probabilities (35), the following values for average weights:

$$\overline{\begin{pmatrix} |< 1|F_{-}|\psi >|^2 \\ |< 2|F_{-}|\psi >|^2 \end{pmatrix}} = x \begin{pmatrix} 0 \\ y \end{pmatrix}_1 + y \begin{pmatrix} x \\ y - \varepsilon \end{pmatrix}. \qquad (38)$$

In the first term, the fluctuation fell on a small packet, which was destroyed as a result. But the weight loss in this case is $x < \varepsilon$, which is not enough. Therefore, the first term (and only it) is subjected to repeated NSF, which with probability equal to one reduces the weight of the only remaining packet to the required value. The result is:

$$\overline{\begin{pmatrix} |< 1|F_{-}^2|\psi >|^2 \\ |< 2|F_{-}^2|\psi >|^2 \end{pmatrix}} = x \cdot 1 \cdot \begin{pmatrix} 0 \\ y - \varepsilon + x \end{pmatrix}_2 + y \begin{pmatrix} x \\ y - \varepsilon \end{pmatrix} = \qquad (39)$$

$$= (1-\varepsilon)\binom{x}{y} + \varepsilon^2 \frac{x}{\varepsilon}\left(1-\frac{x}{\varepsilon}\right)\binom{1}{-1} = (1-\varepsilon)\binom{x}{y} + \binom{\Delta}{-\Delta}; \quad \Delta \le \varepsilon^2/4.$$

Cascade NSF for all possible outcomes ends on this. The comparison shows that (39) is much better consistent with (36) than (38). Thus, the NSF cascade improves the compliance of the model with the SQT here too. For a small package, the relative deviation of the ensemble average weight from the standard does not exceed ε.

$$\frac{\Delta}{x} < \varepsilon \ll 1.$$

For a large packet, the relative deviation is of the order of $\varepsilon^2$.

Now suppose that WF contains two small packages and a number of large packages, which we again combine into one combined package. We get only three packages, the weights of which are again combined into one column vector:

$$\begin{pmatrix} |<1|\psi>|^2 \\ |<2|\psi>|^2 \\ |<3|\psi>|^2 \end{pmatrix} = \begin{pmatrix} x \\ y \\ z \end{pmatrix}, \qquad x+y+z=1, \quad x,y<\varepsilon.$$

After the first NSF averaging gives:

$$\overline{\begin{pmatrix} |<1|F_-|\psi>|^2 \\ |<2|F_-|\psi>|^2 \\ |<3|F_-|\psi>|^2 \end{pmatrix}_1} = x\begin{pmatrix} 0 \\ y \\ z \end{pmatrix}_1 + y\begin{pmatrix} x \\ 0 \\ z \end{pmatrix}_1 + z\begin{pmatrix} x \\ y \\ z-\varepsilon \end{pmatrix}. \qquad (40)$$

In the first two terms, weight loss is insufficient.

If the total weight of two small packets exceeds ε, then to complete the cascade in them it will be enough to conduct another NSF:

$$x\begin{pmatrix} 0 \\ y \\ z \end{pmatrix}_1 \to x\left[\frac{y}{1-x}\begin{pmatrix} 0 \\ y-\varepsilon+x \\ z \end{pmatrix} + \frac{z}{1-x}\begin{pmatrix} 0 \\ y \\ z-\varepsilon+x \end{pmatrix}\right]; \qquad x+y \ge \varepsilon.$$

$$y\begin{pmatrix} x \\ 0 \\ z \end{pmatrix}_1 \to y\left[\frac{x}{1-y}\begin{pmatrix} x-\varepsilon+y \\ 0 \\ z \end{pmatrix} + \frac{z}{1-y}\begin{pmatrix} x \\ 0 \\ z-\varepsilon+y \end{pmatrix}\right]. \qquad (41)$$

After the second rally, the total weight of the three packages in all outcomes is equal to the required value of $1-\varepsilon$. Combining (40) and (41), we get:

$$\overline{\begin{pmatrix} |<1|F_\psi>|^2 \\ |<2|F_\psi>|^2 \\ |<3|F_\psi>|^2 \end{pmatrix}} = (1-\varepsilon)\begin{pmatrix} x \\ y \\ z \end{pmatrix} + \begin{pmatrix} \Delta x \\ \Delta \mathrm{y} \\ -\Delta x - \Delta y \end{pmatrix}; \tag{42}$$

$$\begin{matrix} \Delta x = \varepsilon^2\left[\frac{x}{\varepsilon}\left(1-\frac{x}{\varepsilon}\right) - \frac{\varepsilon}{1-y}\cdot\frac{y}{\varepsilon}\left(1-\frac{y}{\varepsilon}\right)\right] < \varepsilon^2\frac{x}{\varepsilon}\left(1-\frac{x}{\varepsilon}\right) \le \varepsilon^2/4 \\ \Delta y = \varepsilon^2\left[\frac{y}{\varepsilon}\left(1-\frac{y}{\varepsilon}\right) - \frac{\varepsilon}{1-y}\cdot\frac{x}{\varepsilon}\left(1-\frac{x}{\varepsilon}\right)\right] < \varepsilon^2\frac{y}{\varepsilon}\left(1-\frac{y}{\varepsilon}\right) \le \varepsilon^2/4 \end{matrix}; \tag{43}$$

$$\left(\frac{\Delta x}{x}, \frac{\Delta y}{y}\right) < \varepsilon \ll 1, \qquad x + y \ge \varepsilon. \tag{44}$$

If the total weight of two small packets does not reach ε, then to complete the cascade in the first two terms of (40), two more consecutive NSFs must be conducted. Simple calculations similar to (41) give the same result (42), but with slightly different deviations:

$$\begin{matrix} \Delta x = \varepsilon^2(1-y)\frac{x}{\varepsilon(1-y)}\left(1-\frac{x}{\varepsilon(1-y)}\right) \le \varepsilon^2/4 \\ \Delta y = \varepsilon^2(1-x)\frac{y}{\varepsilon(1-x)}\left(1-\frac{y}{\varepsilon(1-x)}\right) \le \varepsilon^2/4 \end{matrix}, \qquad \begin{matrix} x + y < \varepsilon. \\ \left(\frac{\Delta x}{x}, \frac{\Delta y}{y}\right) < \varepsilon \ll 1. \end{matrix} \tag{45}$$

Comparing (39) with (43), (45), we see that with an increase in the number of small packets, the deviation of the average weights from the standard values does not increase, but, conversely, decreases. This is due to the fact that the deviations of small packets from the ideal average weight are always positive. That is, fluctuating small packets are always on average slightly “fuller” than standard ones. If there are a lot of small packages, then they pull off that extra weight from each other, and the agreement with SQT increases. In particular, if WF consists of only small wave packets and all their weights are the same, then after the cascade NSF their average weights will be exactly equal to the standard ones, which immediately follows simply from the symmetry of the problem.

$$\overline{\begin{pmatrix} |<1|F_\psi>|^2 \\ |<2|F_\psi>|^2 \\ \vdots \\ |<n|F_\psi>|^2 \end{pmatrix}} = (1-\varepsilon)\begin{pmatrix} x \\ x \\ \vdots \\ x \end{pmatrix}.$$

If the weights of the packages are different, then the relative deviation of the average weight of each package from the standard will in any case not exceed ε, as in (44).

As for the remaining equalities (37) for average WF values, the cascade here does not give advantages over single NSF. For small packets (as opposed to large ones), the factor $c_{i-}$ is no longer possible to be made constant. It begins to depend on the weight of the package (29):

$$c_{i-} = 1 - x_i \neq 1 - \varepsilon, \qquad x_k < \varepsilon.$$

However, the relative deviation from the ideal is also small here and has the order of $\varepsilon \ll 1$.

The NSF cascade model satisfies all the SQT matching criteria and allows us to close the EQF scheme. However, it is not the only possible scheme for processing wave packets of light weight. In the theory of the full operator of fluctuations (and not the fluctuations operator of norm only, as we have here) there other opportunities arise for combining small amplitudes for the production of NSF, which are not related to the repeated drawings. In these alternative schemes, one can achieve exact SQT matching.

# МОДЕЛЬ ЭНДОГЕННЫХ ФЛУКТУАЦИЙ НОРМЫ ВОЛНОВОЙ ФУНКЦИИ

**Г. А. Птицын**

*Федеральное государственное бюджетное учреждение науки*

*Институт химической физики им. Н.Н. Семёнова*

*Российской академии наук*

*E-mail: gap19542017@gmail.com*

Представлена теоретическая модель эндогенных флуктуаций нормы волновой функции, согласующаяся во всем диапазоне амплитуд со стандартной квантовой теорией. Флуктуации нормы являются подсистемой эндогенных квантовых флуктуаций и описывают один из каналов декогеренции и динамику коллапса квантового состояния.

## ВВЕДЕНИЕ

С развитием квантовых технологий вычислений, связи, «телепортации», шифрования и т.п. на первый план выходят процессы потерь когерентности и информации в сложных квантовых системах, конечным результатом которых является коллапс состояния. В связи с этим разработка теории квантовой декогеренции и коллапса становится все более актуальной.

Стандартная квантовая теория (СКТ) не описывает динамику спонтанной катастрофической перестройки волновой функции (ВФ), происходящей при каждом измерении (а значит и при взаимодействии квантовой системы с любой макроскопической системой). Этот пробел теории отчасти объясняется тем, что проблему «коллапса» удалось обойти.

Несмотря на то, что волновую функцию разрушает именно прибор, он, парадоксальным образом, не влияет на вероятности получения того или иного результата измерения. Эти вероятности полностью определяются видом ВФ системы перед измерением. А именно, вероятность получить при измерении физической величины $A$ значение $a$ определяется выражением:

$$w_A(a) = | < a|\psi > |^2, \quad (1)$$

где $|\psi >$ - вектор состояния системы перед измерением, а $< a|$ - собственный вектор оператора физической величины $A$, отвечающий собственному значению $a$. Этот универсальный закон называется «аксиомой измерения».

Мы выводим декогеренцию и коллапс ВФ из идеи универсальных и неустранимых эндогенных квантовых флуктуаций (ЭКФ), которым, как мы считаем, подвержены все частицы материи [1-4]. Шумит при этом сама ВФ, а общий уровень шума пропорционален числу частиц в физической системе. В малочастичных квантовых системах этот шум крайне мал, и такие системы эволюционируют в согласии с уравнением Шредингера. Однако при измерении такая система запутывается с гигантским числом степеней свободы макроприбора, и шум прибора приводит к практически мгновенному коллапсу объединенной ВФ «частица+прибор».

В картине мира, которую мы предлагаем, нет принципиальной разницы между «квантовой частицей» и «классическим прибором». Здесь все квантовое, и все шумит. Разница не качественная, а количественная – в уровне собственного шума. Малочастичные системы ведут себя как чисто квантовые. Макроскопические, наоборот, находятся в состоянии перманентного коллапса, делающего их макрооднозначными, классическими объектами. Мезоскопические и малочастичные системы на достаточно больших временах демонстрируют постепенную потерю когерентности и информации.

## АКСИОМА ИЗМЕРЕНИЯ

Из сказанного можно сделать важный вывод: для разработки теории декогеренции и коллапса нет нужды моделировать макроприбор и процессы в нем. Достаточно взять одну частицу и подобрать для нее такой алгоритм собственных флуктуаций, который будет удовлетворять аксиоме измерения (1). Такие флуктуации будут постоянно действующим деструктивным фактором в динамике, стремящимся разрушить квантовое состояния и довести его до полного коллапса с правильными вероятностями конечных состояний. В отсутствии прибора частица как бы пытается измерить сама себя, т.е. локализоваться в фазовом пространстве (точнее, в собственном представлении полного оператора флуктуаций [4]). Но это долгий процесс, и можно сказать, что роль прибора в этом случае играет время.

Если такой универсальный алгоритм существует, то он автоматически обеспечит выполнение аксиомы измерения и при взаимодействии частицы с макродетектором. Разница будет лишь в скорости коллапса. Для этого механизм флуктуаций должен обеспечивать кумулятивное действие шумов отдельных частиц на скорость разрушения состояния. Постулируемая универсальность алгоритма ЭКФ сразу объясняет и отмеченную выше независимость вероятностей коллапса от внутренних переменных прибора.

В [4] уже был представлен явный вид полного оператора ЭКФ для одной частицы и алгоритм его действия. Однако там он был недоопределен в области малых амплитуд ВФ. Наша цель - устранить эту недоработку и найти общий вид алгоритмов, удовлетворяющих аксиоме измерения.

Для этого в данном исследовании мы упростим задачу и будем строить не полный оператор флуктуаций, а только его часть, ответственную за уничтожение волновых пакетов и за коллапс состояния. Для этого заметим, что полный оператор флуктуаций действует в фазовом пространстве точечно [4]. При этом он слегка изменяет норму волнового пакета, в котором произошла флуктуация, а также форму ВФ в этом пакете. Эти два действия можно формально разделить, представив оператор флуктуаций $\hat{F}$ в виде произведения двух отдельных (уже нелокальных) взаимно коммутирующих операторов флуктуаций нормы $\hat{F}_n$ и формы $\hat{F}_f$:

$$\hat{F} = \hat{F}_n\hat{F}_f = \hat{F}_f\hat{F}_n.$$

Оператор $\hat{F}_f$ добавляет в ВФ примеси новых состояний, но не меняет норму ВФ, а $\hat{F}_n$, наоборот, меняет лишь норму одного волнового пакета в суперпозиции состояний, совсем не меняя его содержания. Из этого следует, что собственным представлением оператора флуктуаций нормы может служить набор произвольных взаимно ортогональных волновых пакетов в представлении полного оператора флуктуаций, на которые может быть разложена ВФ частицы (что всегда возможно).

Чтобы упростить разработку алгоритма, мы рассмотрим частицу не в произвольном, а в специально приготовленном состоянии, состоящем из $n$ пространственно разделенных взаимно ортогональных волновых пакетов $|\varphi_i(t) >$, и поместим ее в такие условия, чтобы эти пакеты в процессе эволюции не смешивались между собой. В таком случае пакеты и далее будут оставаться ортогональными друг другу, хотя их форма, возможно, и будет меняться. Поскольку нас сейчас не интересует форма пакетов, мы временно заменим реальные эволюционирующие пакеты абстрактными фиксированными ортонормированнми векторами $|i >$, которые будут играть роль суррогатного базиса для $\hat{F}_n$. Суррогатная ВФ имеет вид:

$$|\tilde{\psi} > = a_1|1 > + a_2|2 > + \cdots + a_n|n >, \qquad (2)$$

$$< i|j > = \delta_{ij}, \qquad \sum_{i=1}^{n}|a_i^2| = 1.$$

Если бы собственных флуктуаций не было, то норма каждого пакета $|a_i|$ сохранялась. В нашей модели мы принимаем, что флуктуации волновой функции повторяются через малые равные промежутки времени $\tau$, и квадрат нормы волнового пакета при каждой флуктуации может меняться на малую, но конечную и фиксированную величину $\varepsilon \ll 1$. Таким образом, эволюция истинной ВФ имеет вид:

$$|\psi(t+\tau) > = \hat{F}\hat{U}(\tau)|\psi(t) > = \hat{F}_n\hat{F}_f\hat{U}(\tau)|\psi(t) > ,$$

где $U(\tau)$ – шредингеровский оператор эволюции. Из определения $\hat{F}_n$ следует, что для ВФ, состоящей их пространственно разделенных волновых пакетов, этот оператор будет коммутировать не только с $\hat{F}_f$, но и с $\hat{U}(\tau)$. А это значит, что флуктуации нормы пакетов можно рассматривать совершенно отдельно от эволюции формы самих пакетов. Поэтому мы будем анализировать эволюцию суррогатной ВФ, которая существенно проще:

$$|\tilde{\psi}(m\tau) > = \hat{F}_n{}^m |\tilde{\psi}(0) > ,$$

причем нелинейный случайный оператор $\hat{F}_n$ не меняет базисные вектора $|i >$, а действует, фактически, только на коэффициенты при них. Далее мы будем работать только с суррогатной ВФ и потому опустим знак «~» в ее обозначении. Общий вид истинной ВФ можно восстановить в любой момент времени простой заменой суррогатных базисных векторов $|i >$, на реальные волновые пакеты, изменяющиеся под влиянием операторов эволюции и флуктуаций формы:

$$|i > \xrightarrow{t=m\tau} |\varphi_i(m\tau) > = \left[\hat{F}_f \hat{U}(\tau)\right]^m |\varphi_i(0) > .$$

Рассмотрим один из членов суперпозиции (2) и выясним, каким условиям должны удовлетворять его флуктуационные переходы для того, чтобы его выживание удовлетворяло аксиоме измерения. Будем для краткости называть квадрат нормы пакета его «весом». Если вес пакета равен $x$:

$$|< i|\psi >|^2 = \left|a_i^2\right| = x, \qquad 0 \le x \le 1,$$

то, согласно аксиоме измерения, вероятность $w_\infty(x)$ выживания этого пакета при $t \to \infty$ (в отсутствии макроприбора роль детектора играет время!) должна быть равна $x$:

$$w_\infty(x) = x. \qquad (3)$$

Вероятность коллапса ВФ в данный пакет зависит только от текущего состояния квантовой системы и не зависит от ее предыстории. Это означает, что приводящий к коллапсу случайный процесс является марковским.

Если полный вес всех пакетов в состоянии частицы равен единице, то вероятность выживания всего состояния, согласно (3) также равна единице, т.е. один из пакетов точно выживет:

$$w_\infty = w_\infty(x_1) + w_\infty(x_2) + \cdots w_\infty(x_n) = x_1 + x_2 + \cdots + x_n = 1. \qquad (4)$$

Таким образом, аксиома измерения самосогласована. Но условие самосогласования (4) требует, чтобы вероятность выживания пакета строго соответствовала формуле (3) при любом $x$, в том числе и при сколь угодно малом.

Согласно модели, при каждой флуктуации вес данного пакета может либо уменьшиться $(x \to x - \varepsilon)$, либо остаться неизменным $(x \to x)$, либо увеличиться $(x \to x + \varepsilon)$. Обозначим вероятности этих переходов соответственно $p(x),\ q(x),\ r(x)$. Таким образом, флуктуации веса каждого пакета в отдельности математически представляют собой блуждание на отрезке [0, 1] с поглощающими концами, поскольку точки $x = 0$ и $x = 1$ являются для этого случайного процесса устойчивыми (нулевой вес не может увеличиться, т.к. это означало бы неустойчивость вакуума, аналогично, единичный вес не может уменьшиться, т.к. это означало бы рождение где-то нового пакета с недостающим до единицы весом, т.е. опять неустойчивость вакуума). Для них имеем:

$$w_\infty(0) = 0, \quad p(0) = r(0) = 0, \quad q(0) = 1;$$

$$w_\infty(1) = 1, \quad p(1) = r(1) = 0, \quad q(1) = 1. \qquad (5)$$

Остальные точки отрезка являются неустойчивыми. Полные вероятности ухода из произвольной точки $x$ $(0 < x < 1)$ вниз $(P(x))$, и вверх $(R(x))$ с учетом холостых ходов равны, очевидно:

$$P = p + qp + q^2p + q^3p + \cdots = \frac{p}{1-q} = \frac{p}{p+r}, \qquad (6)$$

$$R = \frac{r}{1-q} = \frac{r}{p+r}, \qquad P(x) + R(x) = 1.$$

Использование вероятностей $P, R$ позволяет избавиться от учета «пропусков» $q$, т.к. последние эффективно содержатся внутри них.

Согласно модели, для точек близких к нулю $0 < x \le \varepsilon$ возможны два флуктуационных перехода: в нуль и в точку $x + \varepsilon$. Для них, исходя из марковости процесса флуктуаций, имеем:

$$w_\infty(x) = P(x)w_\infty(0) + R(x)w_\infty(x + \varepsilon).$$

Подставляя сюда (3), (6), получаем:

$$x = R(x)(x + \varepsilon),$$

т.е.

$$R(x) = \frac{x}{x+\varepsilon}, \;\; P(x) = \frac{\varepsilon}{x+\varepsilon}. \qquad (7)$$

Аналогично, для точек, близких к единице $1 - \varepsilon \le x < 1$ возможны два перехода: в точку $x - \varepsilon$ и в единицу. Соответственно, имеем:

$$w_\infty(x) = P(x)w_\infty(x - \varepsilon) + R(x)w_\infty(1).$$

Используя (3) и (6), получаем:

$$x = \big(1 - R(x)\big)(x - \varepsilon) + R(x) \cdot 1,$$

откуда

$$R(x) = \frac{\varepsilon}{1-x+\varepsilon}, \;\; P(x) = \frac{1-x}{1-x+\varepsilon}. \qquad (8)$$

Вероятности (8) и (7) не симметричны по $P(x)$ и $R(x)$, но симметричны между собой с точностью до одновременной замены $x \leftrightarrow 1 - x$ и $P \leftrightarrow R$.

Для всех остальных точек $\varepsilon \le x \le 1 - \varepsilon$ разрешены симметричные переходы на шаг $\varepsilon$ вверх и вниз. Для них имеем:

$$w_\infty(x) = P(x)w_\infty(x - \varepsilon) + R(x)w_\infty(x + \varepsilon).$$

Подставляя сюда (3) и (6), получаем:

$$x = (1 - R(x))(x - \varepsilon) + R(x)(x + \varepsilon),$$

откуда

$$P(x) = R(x) = 1/2. \qquad (9)$$

Объединяя (5) - (9), получаем окончательно для всех $x$:

$$\begin{cases} p(0) = r(0) = 0, \qquad q\,(0) = 1; & x = 0 \\ p(x) = \dfrac{\varepsilon[1 - q(x)]}{x + \varepsilon},\ r(x) = \dfrac{x[1 - q(x)]}{x + \varepsilon}; & 0 \le x \le \varepsilon \\ p(x) = r(x) = \dfrac{1 - q(x)}{2}; & \varepsilon \le x \le 1 - \varepsilon \\ p(x) = \dfrac{(1 - x)[1 - q(x)]}{1 - x + \varepsilon},\ \ r(x) = \dfrac{\varepsilon[1 - q(x)]}{1 - x + \varepsilon}; & 1 - \varepsilon \le x \le 1 \\ p(1) = r(1) = 0, \qquad q(1) = 1; & x = 1 \end{cases}$$

$$w_\infty(x) = x; \quad 0 \le x \le 1. \tag{10}$$

Построенная модель флуктуационных переходов (10) строго удовлетворяет аксиоме измерения при любом весе пакета. При этом она содержит произвольную пока (кроме конечных точек) функцию $q(x)$ – вероятностей «пропусков». «Пропуски» не влияют на вероятности выживания и уничтожения пакетов. Но их доля, разумеется, будет влиять на кинетику декогеренции квантовых ансамблей и скорость квантового отбора.

В действительности функция $q(x)$ не произвольна, поскольку помимо аксиомы измерения имеется еще несколько критериев соответствия стандартной квантовой теории, которые мы будем обсуждать ниже.

## АДДИТИВНОСТЬ КОЛЛАПСА

Схема (10) не является готовым алгоритмом флуктуаций, поскольку задает вероятности переходов только для одного волнового пакета из полной суперпозиции (2). Чтобы получить алгоритм для всей ВФ, надо учесть связи, которые имеются между отдельными пакетами состояния. А именно, на каждом такте динамики длительностью τ:

- сохраняется полная норма ВФ;
- во всей ВФ происходит ровно одна флуктуация.

Для учета связей надо включить в схему всю ВФ. Сохранение ее нормы означает, что в каждой флуктуации убыль веса в одном пакете компенсируется прибылью в каком-то другом. Т.е. каждая флуктуация содержит два независимых действия - отрицательную и положительную полуфлуктуации (ОПФ и ППФ), а оператор флуктуаций нормы $F_n$ распадается на произведение двух соответствующих операторов $\hat{F}_{n-}$, $\hat{F}_{n+}$.

$$\hat{F}|\psi>= \hat{F}_+ \hat{F}_-|\psi>. \quad (11)$$

(мы опускаем индекс $n$ у этих операторов, т.к. будем заниматься только флуктуациями нормы).

Случайные операторы $\hat{F}_+,\ \hat{F}_-$ не коммутативны. В частности, оператор ОПФ уничтожает пакеты малого веса, после чего ППФ в уничтоженном пакете уже невозможна, в отличие от обратного порядка действий. Мы принимаем порядок, при котором сначала происходит ОПФ, а затем ППФ, как это записано в (11). Важно отметить, что операторы $\hat{F}_+,\ \hat{F}_-$ обозначают внутреннюю нелинейную перестройку ВФ и поэтому являются однонаправленными. В выражениях типа $< a\left|\hat{F}_\pm\right|\psi >$ они действуют только направо и только на кет-вектор состояния. В отличие от линейных операторов физических величин, их действие невозможно перенаправить налево – на вектор представления, каким бы это представление ни было. Более правильной была бы запись: $< a|\hat{F}_\pm\psi >$, но она не всегда удобна. Эрмитово сопряженные к $\hat{F}_+,\ \hat{F}_-$ операторы действуют, наоборот, только налево и только на бра-вектор состояния.

В эксперименте вероятность регистрации частицы пропорциональна сумме весов волновых пакетов, попавших в детектор. Несложно доказать, что для выполнения этого свойства «аддитивности коллапса» достаточно, чтобы были аддитивны (по весу пакетов) алгоритмы каждой из двух полуфлуктуаций. А для этого вероятность полуфлуктуации для двух пакетов, формально рассматриваемых как один комбинированный пакет, должна быть равна сумме вероятностей полуфлуктуаций в каждом из пакетов в отдельности, т.е.:

$$p_\mp(x_1 + x_2) = p_\mp(x_1) + p_\mp(x_2).$$

Это условие удовлетворяется, если вероятности пропорциональны весам пакетов. Нормировать их надо на (текущий) полный вес ВФ:

$$p_{i\mp} = \frac{x_i}{<\psi|\psi>} = \frac{|<i|\psi>|^2}{<\psi|\psi>}. \qquad (12)$$

В этом случае вероятность каждой полуфлуктуации по всей ВФ равна единице, т.е. она где-то точно произойдет, и схема оказывается замкнутой.

Теперь мы построим заново схему переходов для одного волнового пакета с учетом аддитивности полуфлуктуаций.

Вес $x$ волнового пакета уменьшится, если ОПФ произойдет в данном пакете, а ППФ – в каком-то другом. Вес пакета увеличится при противоположной локализации событий. Вероятности $p(x)$, $r(x)$ этих пар независимых событий равны произведению их вероятностей. Согласно (12) это дает:

$$\begin{cases} p(x) = r(x) = \frac{x(1-x)}{(1-\varepsilon)} \\ q(x) = 1 - p(x) - r(x) = 1 - \frac{2x(1-x)}{(1-\varepsilon)} \end{cases}, \qquad \varepsilon \le x \le 1-\varepsilon. \qquad (13)$$

Здесь $(1-\varepsilon)$ – вес всего состояния после ОПФ.

Сравнение показывает, что схема (13) является представителем класса схем (10) в диапазоне весов пакетов ($\varepsilon \le x \le 1-\varepsilon$). Это значит, что в этом диапазоне она строго удовлетворяет аксиоме измерения. Обочины диапазона весов $(0 < x < \varepsilon,\ 1-\varepsilon < x < 1)$ требуют отдельного рассмотрения. Проблема с малыми весами $(x < \varepsilon)$ заключается в том, что при уничтожении такого пакета в результате ОПФ убыль веса состояния не достигает планового значения $\varepsilon$, и это не позволяет удовлетворить всем критериям соответствия СКТ и, в частности, аксиоме измерения. Это значит, что алгоритм флуктуаций для малых пакетов нужно доопределить. Одна из возможностей - это каскадная ОПФ. При этом ОПФ переопределяется не как однократное событие, а как серия последовательных розыгрышей по стандартным правилам ОПФ (12), которые повторяются до тех пор, пока суммарная потеря веса не достигнет точно значения $\varepsilon$. После окончания каскада производится ровно одна ППФ по стандартной схеме с восстановлением веса ВФ до единицы. Обоснование этого рецепта будет дано ниже.

А пока рассмотрим кинетику релаксации статистического ансамбля волновых пакетов, вес которых кратен шагу флуктуации $\varepsilon$: $x = m\varepsilon$ (каждому из этих пакетов соответствуют еще какие-то пакеты, дополняющие полный вес каждого состояния до единицы, но мы сейчас выбираем лишь по одному представителю от каждого состояния).

Согласно схеме, флуктуационные переходы тут идут только по точкам того же набора $x = m\varepsilon$. Запишем текущее состояния этого статистического ансамбля одним вектором-столбцом:

$$|\Phi> = (n_0, n_1, n_2, \dots, n_M)^T .$$

Здесь $n_m$ – доля пакетов в ансамбле с весом $m\varepsilon$; $M = 1/\varepsilon$ – большое целое число; $T$ – символ транспонирования. При такой записи динамику флуктуационных переходов в ансамбле можно описать трехдиагональной статистической матрицей S:

$$|\Phi(t+\tau)> = S|\Phi(t)> ,$$

$$S = \begin{pmatrix} 1 & p_1 & & & \cdots & & & \\ & q_1 & p_2 & & \cdots & & & \\ & r_1 & q_2 & p_3 & \cdots & & & \\ & & r_2 & q_3 & \cdots & & & \\ \vdots & \vdots & \vdots & \vdots & \ddots & \vdots & \vdots & \vdots \\ & & & & \cdots & q_{M-2} & p_{M-1} & \\ & & & & \cdots & r_{M-2} & q_{M-1} & \\ & & & & \cdots & & r_{M-1} & 1 \end{pmatrix}.$$

Здесь $p_m, q_m, r_m \equiv p(m\varepsilon), q(m\varepsilon), r(m\varepsilon)$ из (13). Поскольку $S$ – постоянная матрица, эволюция ансамбля пакетов описывается простым уравнением:

$$|\Phi(n\tau)> = S^n|\Phi(0)>.$$

Решение этого уравнения имеет вид:

$$|\Phi(n\tau)> = \sum_{k=0}^{M} |R_k> C_k {\lambda_k}^n, \qquad C_k = <L_k|\Phi(0)>,$$

$$\lambda_k = 1 - \frac{\varepsilon^2 k(k-1)}{1-\varepsilon}, \qquad k = 0, 1, 2, \dots, M.$$

Здесь $\lambda_k$ – собственные значения матрицы $S$; $<L_k|$, $|R_k>$ - соответственно ее левые и правые собственные вектора со свойствами:

$$<L_k|S = \lambda_k <L_k|, \qquad S|R_k> = \lambda_k|R_k>,$$

$$< R_k|R_k > = 1, \qquad < L_k|R_{k'} > = \delta_{kk'} .$$

Максимальные собственные значения $\lambda_0 = \lambda_1 = 1$ соответствуют двум устойчивым состояниям – нулевому и единичному весам всех пакетов ансамбля. Остальные $(M-1)$ собственных значений распределены в диапазоне [0, 1), обеспечивая релаксацию ансамбля с большим (при $M = \frac{1}{\varepsilon} \gg 1$) набором времен релаксации $T_k = -\tau/ln\lambda_k$. Самое большое из этих времен:

$$T_2 = -\frac{\tau}{ln\lambda_2} \cong \frac{\tau}{2\varepsilon^2} ,$$

можно назвать «временем квантового отбора». За это характерное время эндогенные флуктуации полностью уничтожат все волновые пакеты исходного состояния изолированной частицы кроме одного случайного, который в результате выживет и приобретет максимальный вес за счет погибших собратьев.

Оценки возможных значений параметров модели ε, τ можно будет дать только после обобщения алгоритма флуктуаций на многочастичные системы и рассмотрения модельных процессов декогеренции и измерения.

Построенная модель флуктуационных переходов еще не является готовым алгоритмом. Есть еще критерии соответствия СКТ, которым необходимо удовлетворить перед тем, как начинать компьютерные симуляции флуктуационной квантовой динамики.

## СРЕДНИЕ ЗНАЧЕНИЯ НАБЛЮДАЕМЫХ

Чтобы сформулировать следующие критерии соответствия СКТ нам понадобится явный вид операторов полуфлуктуаций. Согласно схеме, оператор ОПФ действует на случайный волновой пакет $|k>$ в суперпозиции (2) и уменьшает, если это возможно, его вес $x_k$ на фиксированную величину $\varepsilon \ll 1$, либо уничтожает пакет полностью, если его вес не достигает $\varepsilon$. Обозначим потерю веса ВФ при одноразовом розыгрыше ОПФ через $\epsilon$:

$$\epsilon = \begin{cases} \varepsilon, & x_k \geq \varepsilon \\ x_k, & x_k < \varepsilon \end{cases}, \qquad x_k = |< k|\psi > |^2 = |a_k|^2. \quad (14)$$

Тогда для оператора ОПФ получаем следующее выражение:

$$\hat{F}_-|\psi > = (1 + |k > n_k < k|)|\psi >, \quad (15)$$

где $n_k$ – численный множитель, значение которого определяется требуемым снижением веса ВФ в результате ОПФ:

$$< \psi|\hat{F}_-^\dagger \hat{F}_-|\psi > = 1 - \epsilon, \qquad < \psi|\psi > = 1.$$

Подставляя сюда (15), получаем:

$$1 - \epsilon = < \psi|F_-^\dagger F_-|\psi > = < \psi|[1 + |k > (n_k + n_k{}^* + |n_k{}^2|) < k|]|\psi > =$$
$$= < \psi|\ \psi > +(-1 + |1 + n_k|^2)\ | < k|\psi > |^2,$$

т.е.

$$1 - \epsilon = 1 + (-1 + |1 + n_k|^2)x_k. \quad (16)$$

Решая это алгебраическое уравнение с учетом (14), получаем:

$$n_k = -1 + e^{i\xi}\sqrt{1 - \frac{\epsilon}{x_k}} = \begin{cases} -1 + e^{i\xi}\sqrt{1 - \varepsilon/x_k}, & x_k \geq \varepsilon \\ -1, & x_k \leq \varepsilon \end{cases}, \quad (17)$$

где $e^{i\xi}$- произвольный (пока) фазовый множитель.

Положительная полуфлуктуация проще отрицательной. Здесь не бывает каскадов и всегда происходит увеличение веса случайного пакета ровно на $\varepsilon$. Оператор ППФ имеет вид:

$$\hat{F}_+|\psi' > = (1 + |k > p_k < k|)|\psi' >, \quad (18)$$

где $|\psi' > = \hat{F}_-|\psi >$ - ВФ с пониженным весом после ОПФ, а $p_k$ – численный множитель, значение которого определяется требуемым увеличением веса ВФ в результате ППФ:

$$< \psi'|\hat{F}_+^\dagger \hat{F}_+|\psi' > = 1, \qquad < \psi'|\psi' > = 1 - \varepsilon. \quad (19)$$

Подставляя сюда (18) и решая аналогичное (16) алгебраическое уравнение, получаем:

$$p_k = -1 + e^{i\eta}\sqrt{1 + \frac{\varepsilon}{x_k'}}, \qquad x_k' = |< k|\psi' >|^2, \quad (20)$$

где $e^{i\eta}$- произвольный фазовый множитель.

Чтобы найти фазовые множители в (17), (20), надо использовать еще один критерий соответствия модели стандартной квантовой теории. Это требование равенства средних значений физических величин, получаемых во флуктуационной модели и в СКТ, в соответствии со стандартной статистической интерпретацией волновой функции:

$$\bar{A}(t) = Sp[\bar{\varrho}(t)\hat{A}] = Sp[\varrho_S(t)\hat{A}].$$

Здесь $\bar{\varrho}(t)$ – матрица плотности флуктуирующей системы, усредненная по статистическому ансамблю; $\varrho_S(t)$ – матрица плотности СКТ с теми же начальными условиями; $\hat{A}, \bar{A}(t)$ – оператор произвольной наблюдаемой и ее среднее по квантовому ансамблю значение. Решение этого уравнения с произвольным эрмитовым оператором $\hat{A}$ для ВФ (2) приводит к системе равенств для флуктуирующей и стандартной ВФ:

$$\begin{cases} \overline{|< i|\psi(t) > |^2} = |< i|\psi_S(t) >|^2 \\ \overline{< i|\psi(t) >} \ = \ < i|\psi_S(t) > \end{cases}, \quad i = 1, 2, \dots, n. \qquad (21)$$

Чтобы эти равенства выполнялись в произвольный момент времени, они должны выполняться на каждом шаге динамики, а именно перед и после очередной флуктуации. Достаточным условием для этого является выполнение аналогичной пропорциональности для обоих операторов полуфлуктуаций в отдельности:

$$\begin{cases} \overline{\left|< i|\hat{F}_{\mp}|\psi(t) > \right|^2} = k_{\mp}|< i|\psi(t) >|^2 \\ \overline{< i|\hat{F}_{\mp}|\psi(t) >} \ = \ c_{\mp} < i|\psi(t) > \end{cases}, \quad \begin{matrix} i = 1, 2, \dots, n; & (22) \\ k_{\mp},\ c_{\mp} = const. & (23) \end{matrix}$$

### *Отрицательная полуфлуктуация*

Начнем с равенства (22) для ОПФ. До усреднения левой части имеем:

$$\left|< i|\hat{F}_{-}|\psi(t) > \right|^2 = |< i|\,(1 + |k > n_k < k|\,)|\psi >|^2$$

$$= |(1 + \ \delta_{ik}n_i) < i|\psi > |^2$$

Вероятность выпадения ОПФ на $k$-ый пакет равна:

$$w_{k-} = \frac{x_k}{<\psi|\psi>} = x_k = |< i|\psi(t) >|^2. \qquad (24)$$

Усреднение по ансамблю с учетом (17), (24) дает:

$$\overline{\left|< \imath|\hat{F}_{-}|\psi(t) > \right|^2} = (1-\varepsilon)|< i|\psi(t) >|^2, \qquad x_k \geq \varepsilon. \qquad (25)$$

Таким образом, равенство (22) для ОПФ точно выполняется для всех пакетов с весами, превышающими $\varepsilon$. Для константы получаем значение

$$k_{-} = (1-\ \varepsilon).$$

Для пакетов меньшего веса применяется процедура каскадной ОПФ. А именно, если первая ОПФ выпадает на пакет малого веса $x_1 < \varepsilon$, то этот пакет уничтожается, процедура ОПФ повторяется по тем же правилам, но с вероятностями (24), скорректированными на уменьшившийся вес состояния:

$$w_{k-} = x_k/(1-x_1).$$

Эти повторные процедуры с уничтожением пакетов повторяются до тех пор, пока полная убыль веса не достигнет точно значения $\varepsilon$ (т.е. последний пакет в каскаде не уничтожается, а лишь теряет часть своего веса). В приложении к основному тексту мы покажем, что после такого каскада равенство (25) выполняется уже и для малых пакетов (с отклонениями, имеющими следующий порядок малости по шагу флуктуации $\varepsilon$):

$$\overline{\left|< \imath|\hat{F}_{-}|\psi(t) > \right|^2} = (1-\varepsilon)|< i|\psi(t) >|^2 + \Delta_k,$$

$$\Delta_k \leq \varepsilon^2 \frac{x_k}{\varepsilon}\left(1-\frac{x_k}{\varepsilon}\right) \leq \frac{\varepsilon^2}{4}, \qquad x_k < \varepsilon.$$

Кроме того, в теории полного оператора флуктуаций (а не обсуждаемого тут оператора флуктуаций нормы) возникают иные возможности объединения малых амплитуд, не требующие повторных розыгрышей ОПФ. И для них равенство (25) выполняется во всем пространстве уже точно.

Теперь перейдем к равенству (23) для ОПФ. Раскрывая левую часть, получаем:

$$\overline{< \imath\left|\hat{F}_{-}\right|\psi >} = < i\left|\left(1+\left|\overline{k > n_k < k}\right|\right)\right|\psi > = \ (1+\overline{n_\imath}x_i) < i|\psi >.$$

Соединяя это с правой частью (23), получаем:

$$\overline{n_\iota} = \frac{c_- - 1}{x_i} = -\frac{c\prime}{x_i}, \qquad c_- = 1 - c'. \qquad (26)$$

Для больших пакетов $(x_k \geq \varepsilon)$, подставляя сюда значение $n_i$ из (17), имеем:

$$\overline{e^{\iota\xi}} \equiv \theta_- = \frac{1 - c\prime/x_i}{\sqrt{1-\varepsilon/x_i}}, \qquad \varepsilon \leq x_i < 1\,.$$

Поскольку $|\theta_-| \leq 1$, то единственно возможное значение константы $c'$ во всем диапазоне $\varepsilon \leq x_i < 1$ это $c' = \varepsilon$. Окончательно получаем:

$$\theta_- = \overline{e^{\iota\xi}} = \sqrt{1 - \varepsilon/x_i}\,, \qquad c_- = 1 - \varepsilon\,, \qquad \varepsilon \leq x_i < 1\,. \qquad (27)$$

Переписывая (23), получаем:

$$\overline{< \iota\big|\hat{F}_-\big|\psi >} = (1 - \varepsilon) < i|\psi >\,, \qquad \varepsilon \leq x_i < 1\,.$$

Обеспечить выполнение условия (27) можно многими способами. Например, так:

$$e^{i\xi} = \begin{cases} 1, & p = \sqrt{1 - \varepsilon/x_i} \\ \pm i, & q = (1 - p)/2 \end{cases}, \qquad \varepsilon \leq x_i < 1\,. \qquad (28)$$

($p$, $q$ – вероятности событий). Таким образом, для больших пакетов равенство (27) строго выполняется. Функция распределения фазового множителя (28) выбрана произвольно. Никаких критериев кроме (27) для ее выбора пока нет.

Для малых пакетов $(x_k \leq \varepsilon)$, подставляя в (26) значение $n_i = -1$ из (17), имеем:

$$c_{i-} = 1 - x_i, \qquad x_k < \varepsilon. \qquad (29)$$

Таким образом, для малых пакетов равенство (23) слегка нарушается. Относительная величина нарушения имеет порядок $\varepsilon \ll 1$.

### *Положительная полуфлуктуация*

ППФ проще, чем ОПФ. Полный вес состояния после ОПФ равен $(1 - \varepsilon)$, так что ничто не мешает увеличить вес любого пакета ровно на $\varepsilon$.

Найдем константу $k_+$ из (22). Раскрывая левую часть с помощью (18) и усредняя ее с учетом нормировки (19), получаем:

$$\overline{\left|< \iota\left|\hat{F}_+\right|\psi' >\right|^2} = |< i|\psi' > |^2 \sum_{k=1}^{n} w_{k+}|1 + \delta_{ik}p_k|^2 =$$

$$= |< i|\psi' > |^2 \left[1 + (-1 + |1 + p_i|^2)\frac{x_i'}{1-\varepsilon}\right] = \frac{|<i|\psi'>|^2}{1-\varepsilon}. \quad (30)$$

Таким образом,

$$k_+ = 1/(1 - \varepsilon)$$

для всех $i$. Вспоминая теперь (19), что $|\psi' > = \hat{F}_-|\psi >,$ и используя (25), (30), получаем усреднение по обеим полуфлуктуациям:

$$\overline{\left|< \iota\left|\hat{F}_+\hat{F}_-\right|\psi(t) >\right|^2} = \frac{\overline{|<\iota|\hat{F}_-|\psi(t)>|^2}}{1-\varepsilon} = |< i|\psi(t) >|^2 + O(\varepsilon^2)\,.$$

Это равенство справедливо (с точностью $\varepsilon^2/4$) во всем пространстве.

Наконец рассмотрим равенство (23) для ППФ. Раскрывая левую часть, получаем:

$$\overline{< \iota\left|\hat{F}_+\right|\psi' >} = < i\left|\left(1 + \left|\overline{k > p_k < k}\right|\right)\right|\psi' > = (1 + \overline{p_\iota}x'_i) < i|\psi' >.$$

Соединяя это с правой частью (23), получаем:

$$\overline{p_\iota} = \frac{c_+ - 1}{x'_i} = -\frac{c\prime}{x'_i}, \qquad c_+ = 1 - c'.$$

Подставляя значение $p_i$ из (20), получаем:

$$\overline{e^{\iota\eta}} = \theta_+ = \frac{1 + c'/x'_i}{\sqrt{1+\varepsilon/x'_i}}, \qquad 0 < x'_i < 1\,. \quad (31)$$

Поскольку $|\theta_+| \leq 1$, то в этом случае единственно возможное при всех $x'_i$ значение константы это $c' = 0$. Окончательно получаем:

$$\theta_+ = \left(1 + \frac{\varepsilon}{x'_i}\right)^{-1/2}, \qquad c_+ = 1, \qquad 0 < x'_i < 1\,.$$

Обеспечить выполнение условия (31) можно многими способами. Например, так:

$$e^{i\eta} = \begin{cases} 1, & p = 1/\sqrt{1 + \varepsilon/x'_i} \\ \pm i, & q = (1 - p)/2 \end{cases}, \qquad 0 < x'_i < 1\,. \quad (32)$$

Теоретических критериев для выбора распределения этого фазового множителя пока нет. Возможно, какую-то информацию о распределениях

(28) и (32) можно почерпнуть, изучая декогеренцию в интерференционных экспериментах.

Переписывая (23), получаем:

$$\overline{< \imath|\hat{F}_+|\psi' >} = < i|\psi' > , \qquad 0 < x'_i < 1 .$$

Усредняя по обеим полуфлуктуациям, получим:

$$\overline{< \imath|\hat{F}_+\hat{F}_-|\psi >} = (1-\epsilon) < i|\psi > , \quad \epsilon = \min(\varepsilon, x_i),$$

что является наилучшей возможной аппроксимацией желаемого равенства (21), учитывая, что $\epsilon \ll 1$.

Соберем, под конец, полный алгоритм ЭКФ нашей упрощенной флуктуационной модели, возвращаясь к исходным обозначениям. Итак, мы имеем дело с одночастичной ВФ, состоящей из взаимно ортогональных пространственно разнесенных волновых пакетов.

$$|\psi > = a_1|\varphi_1 > + a_2|\varphi_2 > + \cdots + a_n|\varphi_n >,$$

$$< \varphi_i|\varphi_j > = \delta_{ij}, \qquad \sum_{i=1}^{n}|a_i^2| = 1.$$

Через равные микроскопические интервалы времени τ в волновой функции частицы происходит эндогенная флуктуация, описываемая случайным нелинейным оператором $\hat{F}$. В промежутках между флуктуациями ВФ эволюционирует под управлением стандартного шредингеровского оператора эволюции $\hat{U}(t)$. Оператор флуктуаций можно искусственно разбить на произведение операторов флуктуаций нормы $F_n$ и флуктуаций формы $F_f$:

$$\hat{F} = \hat{F}_n\hat{F}_f = \hat{F}_f\hat{F}_n.$$

Эволюция за один динамический цикл τ описывается уравнением:

$$|\psi(t+\tau) > = \hat{F}\hat{U}(\tau)|\psi(t) > = \hat{F}_n\hat{F}_f\hat{U}(\tau)|\psi(t) >,$$

причем для выбранного нами вида ВФ комбинация операторов $\hat{F}_f\hat{U}(t)$ действует только на волновые пакеты $|\varphi_i >$, меняя их форму, а оператор $\hat{F}_n$ действует, фактически, только на коэффициенты $a_i$ при них. Таким образом

динамика расщепляется на два параллельных и независимых процесса: изменения формы волновых пакетов и флуктуации их норм.

Оставляя в стороне изменение формы пакетов, мы построили явный вид только оператора флуктуаций нормы $\hat{F}_n$. Этот оператор распадается на произведение некоммутирующих операторов отрицательной $\hat{F}_{n-}$ и положительной $\hat{F}_{n+}$ полуфлуктуаций.

$\hat{F}_n = \hat{F}_{n+}\hat{F}_{n-}$

Первой происходит отрицательная полуфлуктуация в случайном пакете $|\varphi_k>$ из числа имеющихся с вероятностью

$$w_{k-} = \frac{x_k}{<\psi(t)|\psi(t)>}, \qquad x_k = |< i|\psi(t)>|^2\,, \tag{33}$$

где $<\psi(t)|\psi(t)>$ - полный вес ВФ перед этой ОПФ. В результате ОПФ вес этого пакета уменьшается на величину $\varepsilon$, а если вес пакета не достигает $\varepsilon$, то он уничтожается полностью, а процедура ОПФ повторяется по тому же сценарию, но с вероятностями (33), скорректированными из-за уже слегка изменившегося веса ВФ. Каскад повторов ОПФ продолжается до тех пор, пока полная потеря веса ВФ не достигает точно значения $\varepsilon$. Если пакет в результате ОПФ не уничтожается полностью, то меняется не только его вес, но и комплексная фаза. Явный вид оператора ОПФ таков:

$$\hat{F}_{-}|\psi> = (1 + |k> n_k < k|)|\psi>,$$

$$n_k = \begin{cases} -1 + e^{i\xi}\sqrt{1-\varepsilon'/x_k}, & x_k \geq \varepsilon' \\ -1, & x_k \leq \varepsilon' \end{cases},$$

$$\overline{e^{i\xi}} = \sqrt{1-\varepsilon'/x_k}, \qquad \varepsilon \leq x_k < 1,$$

$$e^{i\xi} = \begin{cases} 1, & p = \sqrt{1-\varepsilon'/x_k} \\ \pm i, & q = (1-p)/2 \end{cases}.$$

Здесь $p$, $q$ – вероятности событий, а $\varepsilon' = \varepsilon$ в случае однократной ОПФ, и $\varepsilon' < \varepsilon$ в случае каскадной ОПФ.

Следом происходит ППФ – тоже в случайном пакете с такой же функционально вероятностью

$$w_{k+} = \frac{x'_k}{< \psi' | \psi' >}. \qquad x'_k = |< i | \psi' >|^2,$$

где $|\psi' >$, $x'_k$ – ВФ и веса пакетов, измененные предшествующей ОПФ. В результате ППФ вес одного пакета увеличивается ровно на $\varepsilon$, тем самым восстанавливая полный вес ВФ до единицы. Кроме веса меняется и его фаза. Явный вид оператора ППФ таков:

$$\hat{F}_+ | \psi' > = (1 + |k > p_k < k|) | \psi' >,$$

$$p_k = -1 + e^{i\eta} \sqrt{1 + \frac{\varepsilon}{x'_k}}, \qquad x'_k = |< k | \psi' >|^2,$$

$$\overline{e^{i\eta}} = 1/\sqrt{1 + \varepsilon / x'_k}, \qquad e^{i\eta} = \begin{cases} 1, & p = 1/\sqrt{1 + \varepsilon / x'_k} \\ \pm i, & q = (1 - p)/2 \end{cases}.$$

## ЗАКЛЮЧЕНИЕ

Разработанный алгоритм флуктуаций нормы войдет составной частью в алгоритм полного оператора флуктуаций. Однако он интересен и сам по себе. Благодаря своей простоте он может применяться для ускорения счета вместо полного алгоритма флуктуаций в тех случаях, когда волновая функция частицы состоит из большого числа волновых пакетов и когда точная форма этих пакетов по какой-то причине не важна. В этом случае вместо реальных волновых пакетов, флуктуирующих по форме и по норме, можно использовать в симуляциях суррогатные пакеты, подчиняющиеся стандартному уравнению Шредингера, т.е. пренебречь флуктуациями их формы и оставить в динамике только флуктуации их нормы.

Развиваемый нами флуктуационный подход к квантовой теории дает обоснование статистической интерпретации волновой функции в стандартной квантовой теории, явно вводя в динамику случайный процесс эндогенных флуктуаций, ответственный за индетерминированный результат квантовых измерений и за потери когерентности и информации в квантовых системах. Однако в отличие от стандартной волновой функции, имеющей математический статус «волны вероятностей», флуктуирующая волновая функция в нашей картине мира имеет физический статус реального шумящего самодействующего поля, но при этом она абсолютно ненаблюдаема. Наблюдаемой является усредненная по квантовому ансамблю функция, совпадающая со стандартной, которую можно измерить методом «квантовой томографии», но не одноразовая реализация случайной волновой функции.

## ПРИЛОЖЕНИЕ. КАСКАДНАЯ ПОЛУФЛУКТУАЦИЯ

Здесь мы проведем анализ алгоритма каскадной отрицательной полуфлуктуации. Мы решили перенести его в конец статьи, чтобы сделать яснее логику основного текста. Однако эта часть алгоритма является одной из главных целей статьи, поскольку она замыкает алгоритм флуктуаций при обращении с волновыми пакетами малого веса, и именно эта часть не была готова в ранее опубликованной версии алгоритма [4].

Согласно схеме, каждая эндогенная флуктуация начинается с отрицательной полуфлуктуации. В супераозиции состояний (2)

$$|\tilde{\psi}> = a_1|1> + a_2|2> + \cdots + a_n|n>, \qquad (34)$$

$$< i|j> = \delta_{ij}, \qquad \sum_{i=1}^{n}|a_i^2| = 1.$$

в результате ОПФ вес одного из волновых пакетов уменьшается на ε. Если же вес пакета меньше ε, то пакет уничтожается полностью. Но тут возникает проблема, потому что симметричные вероятности положительной и отрицательной полуфлуктуаций (12):

$$p_{i\mp} = \frac{x_i}{<\psi|\psi>} = \frac{|<i|\psi>|^2}{<\psi|\psi>}. \qquad (35)$$

диктуемые принципом «аддитивности коллапса», невозможно согласовать с несимметричными вероятностями переходов, диктуемыми для малых пакетов аксиомой измерения (10):

$$p(x) = \frac{\varepsilon[1-q(x)]}{x+\varepsilon}, \qquad r(x) = \frac{x[1-q(x)]}{x+\varepsilon}, \qquad 0 \le x \le \varepsilon.$$

Согласование возможно (и было проведено выше) только для пакетов «тяжелее» ε. Таким образом, алгоритм нуждается в изменении для пакетов малого веса.

Жесткие требования модели таковы:

- норма ВФ должна сохраняться после каждой флуктуации (ОПФ + ППФ);
- малые пакеты должны гибнуть (иначе не будет коллапса).
- коллапс должен быть аддитивным для любых пакетов;
- выживание любого пакета должно соответствовать аксиоме измерения;

Менее жесткое требование: средние значения физических величин, получаемых во флуктуационной модели и в стандартной квантовой теории, должны совпадать.

Чтобы удовлетворить жестким требованиям, потеря веса в ОПФ должна быть доведена до стандартного значения ε. Один из способов этого достичь это каскад повторных ОПФ. После уничтожения пакета с весом $x_1 < \varepsilon$ проводится повторный розыгрыш по тем же правилам ОПФ, но с вероятностями (35), скорректированными на уменьшившийся вес состояния $< \psi|\psi > = 1 - x_1$. Если вторая ОПФ снова выпадает на малый пакет и $x_2 < \varepsilon - x_1$, то этот пакет тоже уничтожается, и проводится третий розыгрыш, и т.д. Последний пакет в серии не уничтожается, а лишь уменьшается в весе так, чтобы суммарная потеря веса равнялась ε. После этого проводится единственная ППФ по стандартным правилам.

Этот алгоритм сохраняет норму ВФ и уничтожает малые пакеты. Он аддитивен по ППФ. Он аддитивен и по ОПФ, так как аддитивен каждый розыгрыш в каскаде. Он удовлетворяет аксиоме измерения, поскольку правила увеличения веса при ППФ не менялись, а ОПФ, как и требуется, уменьшает вес ВФ ровно на ε, просто используя для этого несколько случайных пакетов, если одного окажется мало. В результате каскадного сложения весов малых пакетов алгоритм никогда не попадает в область малых весов, где требования двух жестких принципов не согласуются друг с другом. Таким образом, все жесткие требования модели в каскадной ОПФ соблюдены. Нам остается проверить поведение средних.

Выше было показано, что требование равенства средних значений в модели и СКТ приводят к системе равенств, которые должны выполняться отдельно для ОПФ и ППФ. Алгоритм ППФ не менялся, поэтому полученные там равенства выполнятся и тут. А по отношению к ОПФ они формулируются так:

$$\begin{cases} \overline{|< \imath|\hat{F}_{-}|\psi(t) > |^2} = k_{-}|< i|\psi(t) >|^2 \\ \overline{< \imath|\hat{F}_{-}|\psi(t) >} \quad = c_{-} < i|\psi(t) > \end{cases}, \quad \begin{matrix} i = 1, 2, \dots, n; & (36) \\ k_{-} = c_{-} = 1 - \varepsilon. & (37) \end{matrix}$$

Здесь $\hat{F}_{-}$ - каскадный оператор ОПФ, усреднение идет по квантовому ансамблю, и равенства должны выполняться для каждого каскада.

Если в ВФ есть малые пакеты, то при реализации ОПФ будут возникать каскады. Однако длина каскада сама есть величина случайная. Из-за этого провести расчет средних в общем виде не представляется возможным. Рассмотрим частные случаи. Начнем с равенств (36).

Пусть ВФ состоит из одного малого волнового пакета и некоторого количества больших, которые мы формально объединим в один комбинированный пакет, что облегчит анализ. (Это допустимо благодаря аддитивности схемы флуктуаций по весам. И нас не интересует, как распределяются ОПФ по большим пакетами, т.к. для них алгоритм ОПФ отлажен, и там не может быть никаких сюрпризов.) Запишем веса двух сформированных пакетов в виде одного вектора столбца:

$$\begin{pmatrix} |< 1|\psi >|^2 \\ |< 2|\psi >|^2 \end{pmatrix} = \begin{pmatrix} x \\ y \end{pmatrix}, \qquad x + y = 1, \qquad x < \varepsilon.$$

После первой ОПФ получаем, в соответствии с алгоритмом и выражением для вероятностей (35), следующие значения для средних весов:

$$\overline{\begin{pmatrix} |< 1|F_{-}|\psi >|^2 \\ |< 2|F_{-}|\psi >|^2 \end{pmatrix}} = x \begin{pmatrix} 0 \\ y \end{pmatrix}_1 + y \begin{pmatrix} x \\ y - \varepsilon \end{pmatrix}. \qquad (38)$$

В первом слагаемом флуктуация пришлась на малый пакет, который в результате был уничтожен. Но убыль веса при этом равна $x < \varepsilon$, что недостаточно. Поэтому первое слагаемое (и только оно) подвергается повторной ОПФ, которая с вероятностью единица уменьшает вес единственного оставшегося пакета до требуемого значения. В результате получаем:

$$\overline{\begin{pmatrix}|<1|F_-^2|\psi>|^2\\|<2|F_-^2|\psi>|^2\end{pmatrix}} = x\cdot 1\cdot\begin{pmatrix}0\\y-\varepsilon+x\end{pmatrix}_2 + y\begin{pmatrix}x\\y-\varepsilon\end{pmatrix} = \qquad (39)$$

$$= (1-\varepsilon)\begin{pmatrix}x\\y\end{pmatrix} + \varepsilon^2\frac{x}{\varepsilon}\left(1-\frac{x}{\varepsilon}\right)\begin{pmatrix}1\\-1\end{pmatrix} = (1-\varepsilon)\begin{pmatrix}x\\y\end{pmatrix} + \begin{pmatrix}\Delta\\-\Delta\end{pmatrix};\quad \Delta\le\varepsilon^2/4.$$

На этом каскадная ОПФ для всех возможных исходов заканчивается. Сравнение показывает, что (39) намного лучше согласуется с (36), чем (38). Таким образом, каскад ОПФ и тут улучшает соответствие модели с СКТ. Для малого пакета относительное отклонение среднего по ансамблю веса от стандартного не превышает ε.

$$\frac{\Delta}{x} < \varepsilon \ll 1.$$

Для большого пакета относительное отклонение имеет порядок $\varepsilon^2$.

Пусть теперь ВФ содержит два малых пакета и некоторое количество больших, которые мы опять объединим в один комбинированный пакет. Получаем всего три пакета, веса которых снова объединим в один вектор столбец:

$$\begin{pmatrix}|<1|\psi>|^2\\|<2|\psi>|^2\\|<3|\psi>|^2\end{pmatrix} = \begin{pmatrix}x\\y\\z\end{pmatrix},\qquad x+y+z=1,\quad x,y<\varepsilon.$$

После первой ОПФ усреднение дает:

$$\overline{\begin{pmatrix}|<1|F_-|\psi>|^2\\|<2|F_-|\psi>|^2\\|<3|F_-|\psi>|^2\end{pmatrix}_1} = x\begin{pmatrix}0\\y\\z\end{pmatrix}_1 + y\begin{pmatrix}x\\0\\z\end{pmatrix}_1 + z\begin{pmatrix}x\\y\\z-\varepsilon\end{pmatrix}. \qquad (40)$$

В первых двух слагаемых убыль веса недостаточна.

Если суммарный вес двух малых пакетов превышает ε, то для завершения каскада в них достаточно будет провести еще одну ОПФ:

$$x\begin{pmatrix}0\\y\\z\end{pmatrix}_1 \rightarrow x\left[\frac{y}{1-x}\begin{pmatrix}0\\y-\varepsilon+x\\z\end{pmatrix}+\frac{z}{1-x}\begin{pmatrix}0\\y\\z-\varepsilon+x\end{pmatrix}\right]; \qquad x+y\geq\varepsilon.$$

$$y\begin{pmatrix}x\\0\\z\end{pmatrix}_1 \rightarrow y\left[\frac{x}{1-y}\begin{pmatrix}x-\varepsilon+y\\0\\z\end{pmatrix}+\frac{z}{1-y}\begin{pmatrix}x\\0\\z-\varepsilon+y\end{pmatrix}\right]. \tag{41}$$

После второго розыгрыша суммарный вес трех пакетов во всех исходах равен требуемой величине $1-\varepsilon$. Объединяя (40) и (41), получаем:

$$\overline{\begin{pmatrix}|< 1|F_\psi >|^2\\|< 2|F_\psi >|^2\\|< 3|F_\psi >|^2\end{pmatrix}} = (1-\varepsilon)\begin{pmatrix}x\\y\\z\end{pmatrix}+\begin{pmatrix}\Delta x\\\Delta y\\-\Delta x-\Delta y\end{pmatrix}; \tag{42}$$

$$\begin{matrix}\Delta x=\varepsilon^2\left[\frac{x}{\varepsilon}\left(1-\frac{x}{\varepsilon}\right)-\frac{\varepsilon}{1-y}\cdot\frac{y}{\varepsilon}\left(1-\frac{y}{\varepsilon}\right)\right]<\varepsilon^2\frac{x}{\varepsilon}\left(1-\frac{x}{\varepsilon}\right)\leq\varepsilon^2/4\\ \Delta y=\varepsilon^2\left[\frac{y}{\varepsilon}\left(1-\frac{y}{\varepsilon}\right)-\frac{\varepsilon}{1-y}\cdot\frac{x}{\varepsilon}\left(1-\frac{x}{\varepsilon}\right)\right]<\varepsilon^2\frac{y}{\varepsilon}\left(1-\frac{y}{\varepsilon}\right)\leq\varepsilon^2/4\end{matrix}; \tag{43}$$

$$\left(\frac{\Delta x}{x},\frac{\Delta y}{y}\right)<\varepsilon\ll 1, \qquad x+y\geq\varepsilon. \tag{44}$$

Если же суммарный вес двух малых пакетов не достигает ε, то для завершения каскада в первых двух слагаемых (40) нужно провести еще две последовательные ОПФ. Несложные выкладки, аналогичные (41) дают тот же самый результат (42), но с несколько отличающимися отклонениями:

$$\begin{matrix}\Delta x=\varepsilon^2(1-y)\frac{x}{\varepsilon(1-y)}\left(1-\frac{x}{\varepsilon(1-y)}\right)\leq\varepsilon^2/4\\ \Delta y=\varepsilon^2(1-x)\frac{y}{\varepsilon(1-x)}\left(1-\frac{y}{\varepsilon(1-x)}\right)\leq\varepsilon^2/4\end{matrix}, \qquad \begin{matrix}x+y<\varepsilon.\\ \left(\frac{\Delta x}{x},\frac{\Delta y}{y}\right)<\varepsilon\ll 1.\end{matrix} \tag{45}$$

Сравнивая (39) с (43), (45), видим, что с увеличением числа малых пакетов отклонения средних весов от стандартных значений не увеличивается, а, наоборот, уменьшается. Это связано с тем, что отклонения малых пакетов от идеального среднего веса всегда положительны. Т.е. флуктуирующие малые пакеты всегда в среднем чуть «полнее» стандартных. Если же малых пакетов много, то они оттягивают этот лишний вес друг у друга, и согласие с СКТ увеличивается. В частности, если ВФ состоит из одних только малых волновых пакетов и все их веса одинаковы, то после каскадной ОПФ их средние веса будут в точности равны стандартным, что сразу следует просто из симметрии задачи.

$$\begin{pmatrix} |<1|F_\psi>|^2 \\ |<2|F_\psi>|^2 \\ \vdots \\ |<n|F_\psi>|^2 \end{pmatrix} = (1-\varepsilon)\begin{pmatrix} x \\ x \\ \vdots \\ x \end{pmatrix}.$$

Если же веса пакетов различны, то относительное отклонение среднего веса каждого пакета от стандарта во всяком случае не превысит ε, как в (44).

Что касается оставшихся равенств (37) для средних значений ВФ, то каскад не дает здесь преимуществ перед одинарной ОПФ. Для малых пакетов (в отличие от больших) фактор $c_{i-}$ уже невозможно сделать константой. Он начинает зависеть от веса пакета (29):

$$c_{i-} = 1 - x_i \neq 1 - \varepsilon, \qquad x_k < \varepsilon.$$

Впрочем, относительное отклонение от идеала и тут невелико и имеет порядок $\varepsilon \ll 1$.

Каскадная модель ОПФ удовлетворяет всем критериям соответствия СКТ и позволяет замкнуть схему ЭКФ. Однако она является не единственной возможной схемой обработки волновых пакетов малого веса. В теории полного оператора флуктуаций (а не оператора флуктуаций только нормы, как у нас здесь) возникают иные возможности объединения малых амплитуд для производства ОПФ, не связанные с проведением повторных розыгрышей. В этих альтернативных схемах можно добиться точного соответствия СКТ.

Работа выполнена в рамках государственного задания Министерства науки и высшего образования Российской Федерации (тема 0082–2018–0006, регистрационный номер № АААА-А18-118020890097-1).

# СПИСОК ЛИТЕРАТУРЫ